\documentclass[]{aa}
\usepackage{graphicx, amssymb}

\begin{document}
\input{psfig.sty}
\title{{Extragalactic Globular Clusters in the Near-Infrared:\\
V. IC~4051 and NGC~3311}}

\author {Maren Hempel
        \inst{1,2}
        \and
        Doug Geisler
        \inst{3}
	\and D. W. Hoard
	\inst{4}
	\and
	William E. Harris
	\inst{5}
        }

\offprints {Maren Hempel} 

\institute{European Southern Observatory, Karl-Schwarzschild-Str.~2,
85748 Garching, Germany \\ \email{mhempel@eso.org} \and Michigan State
University, Department of Physics and Astronomy, East Lansing, 48824
MI, USA \\ \email{hempel@pa.msu.edu} \and Departamento de F\'isica,
Facultad de Ciencias F\'isicas y Matem\'aticas, Universidad de
Concepci\'on, Casilla 160-C, Concepci\'on, Chile\\
\email{doug@kukita.cfm.udec.cl} \and Spitzer Science Center,
California Institute of Technology, 220-6, Pasadena CA 91125, USA \\
\email{hoard@ipac.caltech.edu} \and Department of Physics and
Astronomy, McMaster University, Hamilton, Ontario L8S 4M1, Canada \\
\email{harris@physics.mcmasters.ca}}

\date{Received 18 August 2004  ; accepted 28 April 2005}

\abstract{We present the results of combined optical and near-infrared
photometry for the globular cluster systems of the giant ellipticals
IC~4051 and NGC~3311. We use the reduced age-metallicity degeneracy in
$(V-I)$~$vs.$~$(V-H)$ color-color diagrams to derive the cumulative
age distribution within the red sub-population of globular clusters
and to search for age sub-populations. The age distribution is then
compared to the one determined for simulated globular cluster systems
in order to set constraints on the relative age and size of these
globular cluster sub-populations. In both galaxies we find a
significant fraction of globular clusters with ages between 2- 5
Gyr. We also investigate the metallicity distribution in both
systems. Small number statistics prevent us from making any definite
statements concerning NGC~3311, but we find that the derived
metallicity distribution of the IC~4051 clusters strongly depends on
the assumed age distribution. Based on our most likely result that
finds a large number of young/intermediate age clusters ($\sim$2 Gyr)
within the selected globular cluster sample, we find metallicity peaks
at $\sim$ -0.2 for the old clusters and +0.8 for the young
clusters. Only few very metal poor clusters are found. However, the
metallicity distribution within the young/intermediate globular
cluster population is significantly affected by our choice of the
applied Single Stellar Population model. The mean metallicity of the
second generation of globular clusters changes from the above
mentioned and extremely high +0.8 dex to +0.2 dex. Note that the model
dependency becomes less severe with an increasing age of the cluster
population.  }

\authorrunning{Hempel et al.}
\titlerunning{Extragalactic Globular Clusters in the Near Infrared {\tt V}}
\maketitle

\section{Introduction}
\label{introductionnic}
Globular cluster systems (GCSs) are widely used to probe
the formation and evolution of their host galaxies
(e.g. \cite{searle78nic}; \cite{zepf93nic}; \cite{ashman98nic};
\cite{kissler02nic}) although they represent only a small fraction of
the galaxy's total luminosity. Hosting an almost perfect single
stellar population (SSP), globular clusters allow the investigation of
galaxies which, due to their large distances, cannot be resolved into
single stars. The investigation of massive early-type galaxies, not found in
the neighborhood of the Milky Way, in particular suffers from the
large distance drawback. The diffuse light of a galaxy is a composite
of all underlying stellar populations which can not be disentangled,
and populations of several ages and/or metallicities might indeed be
hidden in the integrated light (e.g. \cite{larsen03nic}). Detailed
information on the age structure of the galaxy is therefore very
difficult to obtain from the diffuse integrated light. Globular
clusters, on the other hand, are luminous enough to be observed even
at distances of $\sim$100~Mpc. Their stars share the same age and
metallicity, and especially early-type galaxies can host quite
numerous GCSs (\cite{ashman98nic}). The availability of SSP models
(e.g.~\cite{bruz00nic}; \cite{bruz03nic}; \cite{vaz99nic};
\cite{vaz03nic}; \cite{maraston05nic}) for a wide
range of ages and metallicities, applied to deep photometric or
spectroscopic data of GCSs, allows us to set constraints on their age
and chemical composition (e.g. \cite{goudfrooij01anic};
\cite{puzia02nic}; \cite{puzia04bnic}, \cite{hempel03nic}).

This basic tool allows extragalactic GC research to probe a variety
of important questions. Since the discovery of the bimodal color
distributions in globular cluster systems, their origin has been a
source of ongoing debate, although there is general agreement that
distinct globular cluster populations of different age and/or
metallicity are the origin of this bimodality. When and how these
populations were formed and what can be inferred for the evolution of
the host galaxy are some of the most outstanding questions addressed
today, with crucial implications for galaxy formation
theories. Consequently, much effort has been put into the observation
of GCSs in early-type galaxies, which were classically assumed to have
formed early on and to have evolved passively since then. From
previous surveys, it is known that early-type galaxies are, on the one
hand, well represented by the fundamental plane, indicating basic
similarities, whereas their GCSs can show significant differences; for
example with regards to metallicity (e.g. \cite{geisler96nic}), color
distribution (\cite{gebhardt99nic}), and specific frequency
(\cite{harris81nic}; \cite{harris91nic};
\cite{harris01nic}). Therefore, much caution has to be paid before
claiming a universal formation scenario for early-type galaxies. In
fact, it seems more and more likely that we have to abandon the idea of
a single formation theory and find external parameters which might be
decisive for the particular formation and evolution of a given
galaxy. Factors involving the galaxy environment, such as its position within
the host galaxy cluster/group or its isolation from other galaxies, is
likely to be one of the most important external parameters. Galaxy
interactions (e.g. mergers and accretion events) seem to be an
important, if not the dominant,  process, governing galaxy formation and
evolution (see contributions in
\cite{wielen90nic}). Consequently, the GCSs hosted by galaxies which were likely
involved in such interactions need to be investigated. First results
on the investigation of early-type galaxies in low-density
environments were published by Kuntschner et al. (2002), who used
integrated-light spectroscopy to derive ages and metallicities of 40
E and S0 galaxies and conclude that early-type galaxies in a low
density environment show a broad distribution of luminosity-weighted
ages and are on average younger than cluster
ellipticals. Unfortunately, no early-type galaxies in a high galaxy
density environment exist in the very local Universe. The determination of
globular cluster ages by integrated-light spectroscopy is therefore
quite challenging due to the faintness of the targets. The capability
of high signal-to-noise spectroscopy with respect to age and
metallicity determination has been demonstrated by various studies
(e.g. \cite{puzia04anic}; \cite{puzia04bnic}) and spectroscopy will
doubtless be the method of choice when precise age estimates of a
subsample of bright clusters are required. However, as long as the
objective is to detect different broad age subpopulations in GCSs and
to set coarse age limits on the GCS as a whole, then photometric studies
(e.g. \cite{kundu01anic}; \cite{kundu01bnic}; \cite{gebhardt99nic})
are the most efficient alternative. \\

Until recently, the major drawback for the use of photometric studies
to investigate the age distribution in GCSs was the age-metallicity
degeneracy (\cite{worthey94nic}). With the introduction of combined
optical and near-infrared photometry (e.g. \cite{minniti96nic};
\cite{goudfrooij01bnic}; \cite{kissler02nic}) it is now possible to at
least partially overcome this obstacle and derive the relative ages of
globular cluster sub-populations, as well as their metallicities. The
extended wavelength range allows one to recognize intermediate age
globular clusters (e.g. those formed in a recent merger) from the bulk
of first generation objects, formed shortly after the Big Bang. In
addition, colors like V-H provide a sensitive metallicity index. The
integrated IR colors for Galactic globular clusters of Cohen \&
Matthews (1994) show that the $V-H$ index has three times the
metallicity sensitivity of $V-I$, making it a good index for
investigating the reality of any multiple metallicity populations.

\begin{table*}[!ht]
\centering
\caption[General informations about NGC~3311 and IC~4051]{General
information about the host galaxy.}
\vspace{5mm}
\label{parametersnic}
\begin{tabular}{l r r l}
\hline
\noalign{\smallskip}
Property & NGC~3311 & IC~4051 & Reference\\
\noalign{\smallskip}
\hline
\noalign{\smallskip}
RA(J2000)           & 10h 36m 43s            & 13h 00m 55s            &\cite{rc3nic}\\
DEC(J2000)          & $-27^{\rm o}$ 31' 42'' & $+28^{\rm o}$ 00' 27'' &\cite{rc3nic}\\
$B_{\rm T,0}$       & 12.65                  & 14.17                  &\cite{rc3nic}\\
E$_{B-V}$           & 0.079                  & 0.011                  &\cite{schlegel98nic}\\
A$_{V}$             & 0.263                  & 0.035                  &\cite{schlegel98nic}\\
A$_{I}$             & 0.154                  & 0.02                   &\cite{schlegel98nic}\\
A$_{H}$             & 0.046                  & 0.006                  &\cite{schlegel98nic}\\
$(B-V)_{eff}$       & 1.04                   & 1.01                   &\cite{rc3nic}\\
$(m-M)_V$           & $33.70\pm0.08$         & 34.82$\pm0.13$         &\cite{jensen01nic}\\
M$_V$               & $-22.07\pm0.20$    & $-21.62\pm0.35$            &\cite{rc3nic}\\
\noalign{\smallskip}
\hline
\end{tabular}
\end{table*}

In this study we investigate the GCSs of two important early-type galaxies,
viz. NGC~3311, representing a giant cD galaxy at the center of a high
density environment (the Hydra I galaxy cluster), and IC~4051, a giant
E2 in the Coma galaxy cluster with no luminous neighbor, located
several cluster core radii from the center. This will provide a range
of environmental densities in order to begin to probe their effect on the GCS.
The GCS of NGC~3311 has been investigated in the optical by many
different groups (e.g. \cite{grillmair94nic}; \cite{mclaughlin95nic};
\cite{brodie00nic}; \cite{secker95nic}). They found a populous GCS
which appears to show the general color bimodality of such systems.
Due to its central position in a high density environment, several
wide field surveys have been carried out (e.g. \cite{harris83nic};
\cite{hilker03anic}; \cite{hilker03bnic}) to investigate spatial
features in the GCS. NGC~3311 is of special interest in globular
cluster studies, since it is not only the third closest cD galaxy,
following in distance M~87 in Virgo and NGC~1399 in Fornax, but also
hosts one of the richest GCSs known (e.g. \cite{harris83nic};
\cite{mclaughlin95nic}). The GCS of IC 4051 has only been attainable
until now from HST imaging and was the subject of papers by Baum et
al. (1997) and Woodworth and Harris (2000). The latter work used V,~I
photometry and found tentative evidence of bimodality, with
metal-intermediate ([Fe/H]$\sim -0.7$) and metal-rich ($\sim$ solar)
peaks and a near-complete lack of metal-poor clusters. The lack of
metal-poor clusters and the very high specific frequency
($S_N=11\pm2$) that they found makes the GCS of IC 4051 very unique
and worthy of further study.  General information concerning these 2
galaxies can be found in Table 1.

In this paper, we combine optical and near-infrared photometry to
derive the cumulative age distribution (e.g. \cite{hempel04anic})
which we will compare to simulated GCSs of known age structure (size
and age of sub-populations). In this particular case the model
isochrones by Bruzual \& Charlot (2003) have been applied and the
results are discussed in great detail. Nevertheless, an additional
section will be spent to discuss the dependency of the results on the
choice of SSP models. To do so we will repeat the analysis and apply
the latest version of the SSP models by Maraston (2005). These results
enable us to set constraints on the age structure in both systems and
to infer clues to the evolutionary history of their host galaxies
(\cite{hempel04anic}). The paper is organized as follows. In Section
\ref{s:observationnic} we describe the observations as well as the
data reduction. The derivation of the cumulative age distribution
within the observed samples is described briefly in
Sect.~\ref{agenic}. For a detailed description we refer to Hempel et
al. (2003) and Hempel \& Kissler-Patig (2004). In Section
\ref{s:metalnic} we derive and discuss the metallicity distribution.
We summarize our results in Section \ref{s:summarynic}. We will
dedicate Appendix \ref{appendixa} to the discussion of how the results
are influenced by our choice of a Single Stellar Population model.

\vskip 0.5cm
\section{Observations and Data Reduction} 
\label{s:observationnic}

The optical and the near-infrared data for both galaxies were reduced
by different groups, using different software. Therefore
we will give a brief description for each data set.

\begin{table}[!ht]
\centering
\caption[Exposure times]{Exposure times for NGC~3311 and IC~4051}
\vspace{5mm}
\label{exposurenic}
\begin{tabular}{l r r r r}
\hline
\noalign{\smallskip}
Filter (Instrument)  &  &  NGC~3311 &  & IC~4051\\
\noalign{\smallskip}
\hline
\noalign{\smallskip}
F814W	(WFPC2)	    &      & 3800 sec   &    &  5200 sec  \\
F555W   (WFPC2)     &      & 3700 sec   &    &   --       \\
F606W   (WFPC2)     &      & --         &    & 20500 sec  \\
F160W   (NICMOS2)   &      & 2560 sec   &    &  2560 sec  \\
\noalign{\smallskip}
\hline
\end{tabular}
\end{table}

\subsection{Optical data: HST/WFPC2}
\label{s:wfpc2nic}

Both targets were observed with HST + WFPC2 in two optical pass bands,
NGC~3311 in F814W and F555W, and IC~4051 in F814W and F606W,
respectively. NGC~3311 observations are part of a large survey on extragalactic
GCSs, carried out under program number
GO.6554. The observations of IC~4051 were carried out under program
GO.6283, dedicated to stellar populations in elliptical galaxies. The
total exposure times in the various pass bands are given in Table
\ref{exposurenic} for both targets. 

\begin{figure}[!h]
\center 
\includegraphics[width=7cm]{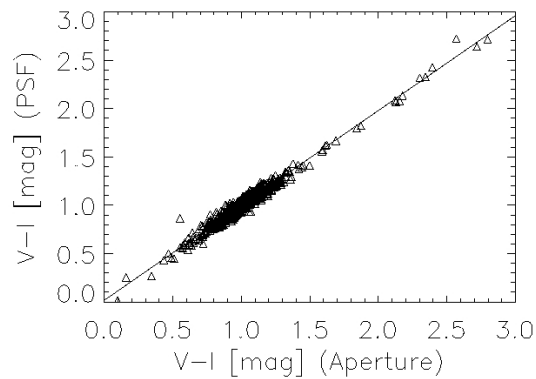}
\caption[Comparison between {\it{psf}} and {\it{aperture}}
photometry]{Comparison between $(V-I)$ colors based on {\it{aperture}}
and {\it{psf}} photometry. The solid line represents the least-squares
fit to $(V-I)$ based on aperture photometry. The objects in this plot
contain the $(V-I)$ colors for the PC1 chip as well as for all three
wide field chips. As mentioned in Sect. \ref{s:observationnic} the
latter could not be included in our age analysis, due to the smaller
spatial coverage of the near-infrared observations.}
\label{comparenic}
\end{figure}

For IC~4051 we took the photometry directly from Woodworth \& Harris
(2000). The reduction of the optical data for NGC~3311 is described in
the following.\\ The photometry on NGC~3311 clusters was done using
the DAOPHOT package for {\it{aperture}}- and alternatively for
{\it{psf}} photometry as well as using SExtractor (\cite{bertin96nic})
{\it{aperture}} photometry. In Figure \ref{comparenic} we show the
comparison between $(V-I)$, calculated from instrumental V and I band
magnitudes, using DAOPHOT {\it{psf}}-photometry and SExtractor
{\it{aperture}}-photometry for all objects detected on the PC1 chip,
for which later on also H-band data were obtained. The extremely red
objects ($(V-I)$$>1.6$) have been rejected for being objects either in
the vicinity of the central dust shell found in the galaxy core or
close to the edge of the PC chip. Including an error selection,
i.e. $\delta$$(V-I)$$<$0.15 mag, the results of a least square fit
between the $(V-I)$ results are:

\begin{equation}
\hspace{1cm} (V-I)_{aperture}=0.983(V-I)_{psf}+0.015
\label{equation1nic}
\end{equation}

There is no difference between the results, the color offset being smaller
than the photometric error, except for very red objects
($(V-I)$$>$~2.0 mag). The NGC~3311 results presented in this Paper are
based on SExtractor photometry. 

The instrumental magnitudes were transformed to the Johnson V and
Cousin I magnitudes according to the procedure given by
Holtzman et al. (1995) and reddening corrected using the values given
in Table \ref{parametersnic}. The aperture corrections for both filter bands were
determined as the mean difference (21 objects) between a 6 pixel
instrumental magnitude and the 0\farcs5 aperture used by Holtzman et al.
(1995). The calculated corrections, 0.229 mag and 0.265 mag for the V
and I band, respectively, were than applied to all of the V- and
I-band detected objects. The small aperture was necessary because of
crowding.\\

The near-infrared data used in this paper were derived from the
instrumental magnitudes obtained in the non-standard filter systems of
NICMOS, using the information of the mean metallicity based on optical
colors. The relation between $(V-I)$ and the metallicity is rather
well calibrated, using the globular clusters of the Milky Way or M31,
respectively. However, considering the fact that the correction
between F160W and the Cousin H-band for a mean metallicity of
$\langle$Fe/H$\rangle$~$\sim$-0.3 (IC~4051) and
$\langle$Fe/H$\rangle$~$\sim$-0.75 (NGC~3311), as described in
Sect. \ref{ic4051dnic} and \ref{n3311dnic} differ only insignificantly
(0.115 mag in IC~4051 compared to 0.110 mag in NGC~3311) leads us to
the conclusion that the a priori assumption of the metallicity does
not alter the final outcome of our analysis. Nevertheless, since
additional near-infrared data are desirable for an improved GC sample
in {\bf{both}} galaxies, observations using broad band filters with a
response function closer to the Johnson-Cousin system are clearly
preferable, or alternatively observations in multiple bands, allowing
a proper color correction.

\vskip 0.2cm
\subsection{Near-Infrared Data: HST/NICMOS2}
\label{s:nicmosnic}
Both galaxies were observed with NICMOS Camera 2 (fov:
19\farcs2$\times$19\farcs2, 0\farcs075 pix$^{-1}$) through the F160W
filter (approximately H band) in a slightly overlapping ($\sim$10-20
pixels) four-point mosaic pattern, centered on the galaxy. The
combined mosaic image covers approximately the same area and galaxy
field as the WFPC2 PC1 chip. Each of the four quadrants is a
combination of 5 sub-images. The exposure time for each sub-image is
512 sec, the total exposure time sums up to 2560 sec (see Table
\ref{exposurenic}).  As for the optical filter bands, we give the
details for data reduction separately for each target.\\

\vskip 0.2cm
\subsubsection{IC~4051 }
\label{ic4051dnic}
In IC~4051 the SExtractor routine (\cite{bertin96nic}) was used to
identify all targets in the four quadrant images of the
galaxy-subtracted NICMOS2 image. This resulted in the detection of a
total of 642 targets. The aperture photometry was then carried out
using the non-standard IRAF task CCDCAP (\cite{mighell95nic}), which
is optimized for aperture photometry of undersampled
images. Hereafter, the photometry procedure recommended in the NICMOS
Data Handbook \footnote{See http:www.stsci.edu/hst/nicmos/}
(\cite{dickinson02nic}) was applied. The instrumental magnitudes were
measured with an aperture of 0\farcs5 (radius, corresponding to 6.6
pixel) and corrected to ``infinite'' aperture radius by adding -0.1517
mag, calibrated by adding 23.566 mag (which is the magnitude of a star
with count rate of 1~electron per second in the NICMOS F160W filter),
and corrected for the total quadrant image exposure time.  Finally, we
performed a transformation from the NICMOS F160W filter magnitudes to
standard H magnitudes (\cite{stephens00nic}). Based on the mean
metallicity of IC~4051 globular clusters of
$\langle$Fe/H$\rangle$~$\sim$-0.1 (\cite{woodworth00nic}), combined
with observations of Galactic and M31 globular clusters
(\cite{barmby00nic}), leads to the following transformation:

\begin{equation}
\hspace{1cm} H= F160W - 0.115~(\pm0.048)~mag
\label{equation2nic}
\end{equation}

After all calibration corrections were applied to the instrumental
magnitudes, a number of targets were rejected: all targets within 12
pixels of any edge of a quadrant image, all targets for which CCDCAP
failed to measure a magnitude, all targets with SExtractor star/galaxy
indices less than 0.5 (indicating they are likely galaxies), targets
located near the the center of the subtracted galaxy (within
$\sim$2\farcs0, where photometric errors are very large due to the
background galaxy light), and for identical targets located within the
overlapping regions of two quadrant images, the one with the larger
uncertainty. A total of 348 objects remained. After transforming the
WFPC2 pixel coordinates to the NICMOS coordinates, and allowing an
offset of up to 2 pixels in both x and y coordinates, the optical and
near-infrared data were matched. The final sample now includes 256
objects.\\

\vskip 0.2cm
\subsubsection{NGC~3311}
\label{n3311dnic}
As in the optical filter bands the source detection on the NICMOS
images was done using DAOPHOT/PHOT as well as SExtractor. To be as
consistent in the data reduction as possible, we will stay with the
SExtractor results. The transformation of the F160W data to the
Johnson H band followed the procedure described in Section
\ref{ic4051dnic} and results in the following correction:
\begin{equation}
\hspace{1cm} H= F160W - 0.110~(\pm0.024)~mag
\label{equation3nic}
\end{equation}

The differences in the transformation between F160W and H-band
magnitudes between NGC~3311 and IC~4051 originates in the difference
of the mean metallicity, derived from the $(V-I)$ distribution, where
we have adopted a mean metallicity of
$\langle$~Fe/H~$\rangle$$\sim$-0.75 (\cite{brodie00nic}). Besides the
F160W- H band transformation, the instrumental magnitudes were
aperture corrected, the photometric zero point for NICMOS2 applied and
finally corrected for galactic reddening (see Table
\ref{parametersnic}). After combining optical and near-infrared data
the globular cluster sample includes 148 objects.\\

In this series of papers we concentrate on the age structure in
globular cluster systems, which we derive from color-color
diagrams. The basic tool in this approach (see Section \ref{agenic}
and \cite{hempel04anic}) is the color distribution of the observed
globular clusters with respect to SSP model isochrones. Therefore we
have to consider the contamination of the globular cluster
samples. Considering the color range, the contaminants are most likely
background galaxies. Although we apply selection criteria, e.g. shape
parameters and color limits, we are aware that contamination is still
an issue. Nevertheless, the final globular cluster samples, combining
optical and near-infrared data, contain only objects in the very
central region of the GCS, where we find the highest globular cluster
density. In combination with the sample size ($\approx$100
objects), as discussed in Section \ref{agenic}, we estimate that
unresolved background galaxies make only a minor contribution to our
sample and should have a minimal effect on the age distribution.

\vskip 0.2cm
\section{Results}
\label{resultsnic}
\subsection{Color-Color Distributions}
\label{colornic}

Our basic tools to derive ages and metallicities of stellar
populations from photometric data are color-color diagrams together
with various SSP models; for example by Bruzual \& Charlot (2000,
2003), Vazdekis (1999), Maraston (2001) or their most recent release
(\cite{maraston05nic}). We will use the Bruzual \& Charlot models
(\cite{bruz03nic}) throughout the main body of the paper whereas
Appendix \ref{appendixa} is based on the Maraston models
(\cite{maraston05nic}) which indicates that only minor differences
occur between models for ages $>$1 Gyr and that our main goal of
detecting different aged subgroups is not compromised. In our further
analysis we select only clusters which satisfy an error limit of
$\delta_{(V-I)}$ and $\delta_{(V-H)}$~$<$ 0.15 mag. The error limit
was derived as a combination of the photometric error in two filter
bands with $\delta$$\leq$0.1mag each, corresponding to a S/N=10 for
detected objects. The error selected globular cluster samples for
NGC~3311 and IC~4051 contain 105 and 149 objects, respectively. In our
further analysis of the age structure additional color cuts will be
applied to minimize contamination with unresolved background objects
and to select objects within a restricted color range that allows the
best age resolution (see Section \ref{agenic}). With respect to
previous studies, e.g. by Hilker (NGC~3311, \cite{hilker02nic},
\cite{hilker03anic}) and Woodworth \& Harris (IC~4051,
\cite{woodworth00nic}) we note the much smaller size of our sample,
which is mostly due to the included H-band data, and the small field
of view of the NICMOS2 instrument. As we will discuss later (see
Sect. \ref{agenic} and \ref{s:summarynic}), the limited spatial
distribution of the globular cluster sample becomes of importance for
our interpretation of the derived relative sizes of the age
sub-populations.  \\

\subsubsection{IC~4051}
\label{resultic4051nic}

\begin{figure*}[!ht]
\center
\includegraphics[width=8cm]{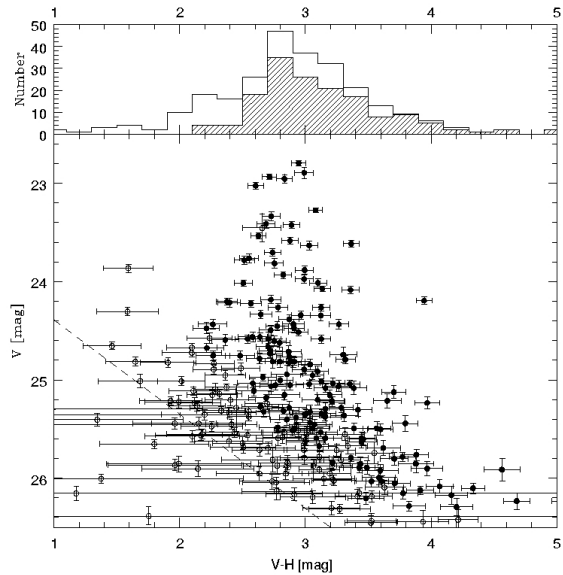}
\caption[$V~vs.~(V-H)$ color-magnitude diagram for IC~4051]{$V~vs.~(V-H)$ color-magnitude diagram for IC~4051. The top
sub-panel shows the color distribution of all (open histogram) and
error-selected (hashed histogram) objects. The lower sub-panel shows
the color-magnitude diagram. Herein the filled symbols mark the
selected clusters, while the open circles mark rejected objects.
The dashed line marks the 50$\%$
completeness limit for the H-band. }
\label{ic4051f4nic}
\end{figure*}

\begin{figure*}[!ht]
\center
\includegraphics[width=8cm]{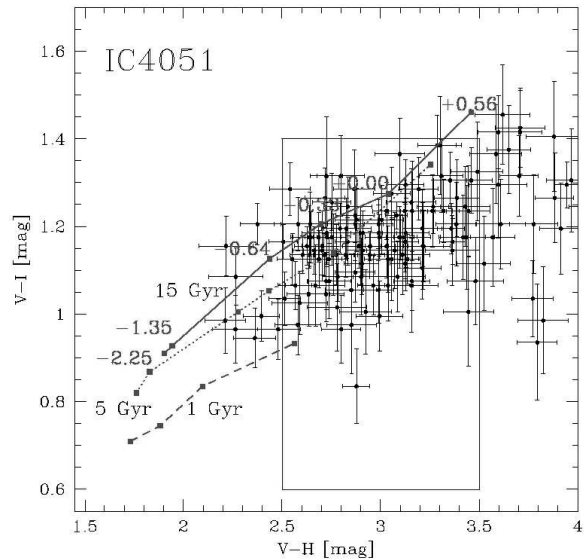}
\caption[$(V-I)$~$vs.$~$(V-H)$ color-color diagram for
IC~4051]{$(V-I)$~$vs.$~$(V-H)$ color-color diagram for IC~4051. All
data are corrected for foreground reddening (see
Table~\ref{parametersnic}).  The 15 Gyr, 5 Gyr and 1 Gyr isochrones
are based on the Bruzual \& Charlot SSP models (\cite{bruz03nic}). The
solid squares mark the metallicity in [Fe/H], increasing with $(V-H)$
from -2.25 up to +0.56 . The box marks the color range selected for the
determination of the cumulative age distribution.}
\label{ic4051f6nic}
\end{figure*}

\begin{figure}[!ht]
\center
\includegraphics[width=9cm]{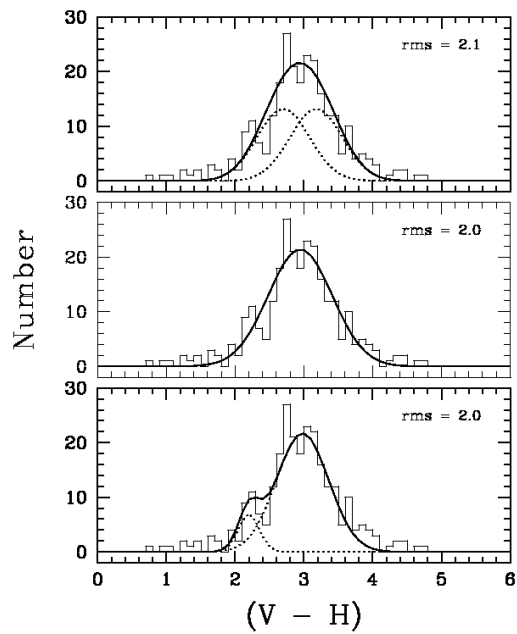}
\caption[Gaussian fit to the $(V-H)$ color distribution in
IC~4051]{$(V-H)$ color distribution in the complete IC~4051 sample
showing a single Gaussian fit (middle) and two different double
Gaussian fits (top: equal amplitude/equal width, bottom: variable
amplitude/variable width). The dotted lines mark the contribution of
each individual component, solid lines are their sum.}
\label{ic4051cmdcompletenic}
\end{figure}

In Figures \ref{ic4051f4nic} and \ref{ic4051f6nic} we show the
color-magnitude diagram (CMD) and the $(V-I)$$~vs.~(V-H)$ color-color
diagram for IC~4051. Clearly, the data below the 50$\%$ completness
limit of the H photometry are unreliable and will be excluded from
further analysis, as will all objects with color errors above our
error cut. In Woodworth \& Harris (2000) it was tentatively suggested
that the $(V-I)$ distribution of the IC~4051 globular clusters is
bimodal and, therefore, we attempted to fit a double Gaussian function
to the $(V-H)$-color distribution, which is three times more sensitive
to metallicity than $(V-I)$. If the complete sample set is included in
this analysis, then comparably good fits are obtained for variable
amplitudes for both sub-populations or (forced) equal amplitude, equal
width settings. In the latter case the Gaussian centers are at
$(V-H)$=2.70 and 3.18. The mean of these two Gaussian centers is 2.94,
which is approximately equal to both the mean of all colors and the
center of the single Gaussian fit. Thus, this fit essentially devolves
to merely reproducing the single Gaussian case. However, since uni-
and bimodal fits give similarly good results (see Figure
\ref{ic4051cmdcompletenic}), we will stay with the assumption of a
unimodal $(V-H)$ color distribution and a corresponding mean color of
$\langle V-H \rangle = 2.92$. We note that the situation changes
drastically if we apply the selection by error limit. As we can see in
Figure \ref{ic4051f4nic}, the blue population ($V-H<$2.4) is
significantly diminished by the error limit, which makes it difficult
to draw strong conclusions from the color distribution. The
metallicity distribution is further investigated in Section
\ref{s:metalnic}.\\

\begin{figure}[!ht]
\center
\includegraphics[width=8cm]{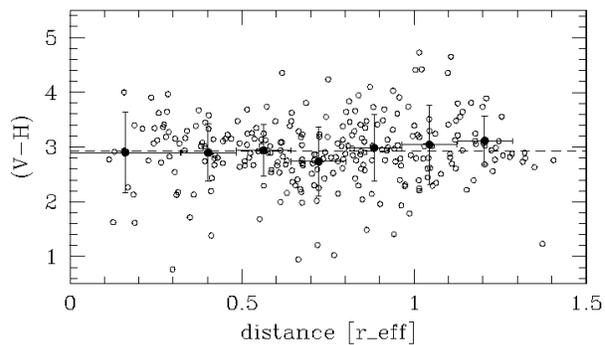}
\caption[Radial distribution of the IC~4051 globular clusters]
{Distribution of globular cluster colors for IC~4051 as a
function of radius from the center of the galaxy (open circles). The
dashed line marks the mean value of the entire data set,
$\langle$V-H$\rangle$=2.922 mag. The solid circles show the mean
colors in increasing radius bins, as listed in Table
\ref{radialdisnic}. }
\label{radialnic}
\end{figure}

With the discovery of bimodal color distributions in globular cluster
systems (e.g. \cite{whitmore95nic}; Geisler, Lee \& Kim 1996;
Kundu \& Whitmore 2001a; Gebhardt \& Kissler-Patig 1999) the interest in the GCSs of early-type
galaxies, which were supposedly old and uniform, grew immensely. The
spatial distribution of GCs adds an extra clue to their origin. For example,
in the Galaxy, metal rich GCs are found to be more confined to the
Galactic disk out to distances up to $\sim$10 kpc from the Galactic
center (\cite{searle78nic}), and metal-poor clusters
are found at even larger distances but spherically distributed in the Galactic
halo (\cite{ashman98nic}). Although the near-infrared data for IC~4051
cover only the innermost region of IC~4051 (maximum distance $\sim$9
kpc) it is still worth investigating the radial distribution of the
globular clusters mean colors. Figure
\ref{radialnic} shows the distribution of globular cluster colors as a
function of galactocentric angular distance R; we have listed the mean
colors in 2\farcs5 wide bins in Table \ref{radialdisnic}. We note that
there is no obvious trend in the mean colors. $V-H$ is primarily
sensitive to metallicity, so this indicates the lack of any strong
metallicity gradient in the IC~4051 GCS, for radial distances R$\leq$22\farcs5.


\begin{table}
\centering
\caption[width=\textwidth]{Mean Color versus Radius for IC 4051
\label{radialdisnic}}
\begin{tabular}{r r c c}
\hline
\noalign{\smallskip}
Radius (\arcsec) & $N$ & $\langle V-H \rangle$ & $\sigma_{V-H}$\\
\noalign{\smallskip}
\hline
\noalign{\smallskip}
 $<$5.0      & 30 & 2.901 & 0.745 \\
 5.0--7.5    & 28 & 2.897 & 0.530 \\
 7.5--10.0   & 41 & 2.938 & 0.478 \\
 10.0--12.5  & 51 & 2.737 & 0.635 \\
 12.5--15.0  & 51 & 2.984 & 0.612 \\
 15.0--17.5  & 38 & 3.044 & 0.724 \\
 17.5--20.0  & 25 & 3.114 & 0.451 \\
\noalign{\smallskip}
\hline
\end{tabular}
\end{table}

Comparing the combined optical and near-infrared colors with SSP model
isochrones (see Figure \ref{ic4051f6nic}) reveals that only a fraction
of the red cluster population is consistent with an old ($>$10 Gyr)
population. We find a large fraction of objects whose $(V-I)$ color at
a given $(V-H)$ is too blue to fit into an ``old'' globular cluster
scenario. Their color is much better matched by an intermediate age
population (i.e. with an age between $\sim$1 and 7 Gyr). Such a large
intermediate/young population of globular clusters seems suspicious,
since the integrated galaxy light does not show any hint of a recent
major star formation event and it is a relatively isolated
galaxy. With respect to the integrated light argument we refer to the
work by Puzia et al. (2002) on NGC~4365 who found a similar scenario
where the combined $(V-I)$ and $(V-K)$ color of the host galaxy is
consistent with a metal-rich and relatively old stellar population
although the globular cluster system indicates a significant fraction
of intermediate age globular clusters as well as old ones
(\cite{puzia02nic}, \cite{hempel04anic}). The apparent contradiction
between the two results breaks down if we compare the observed
fraction of globular clusters with the stellar content of the
galaxy. The derived ratios between age sub-populations in both
galaxies refers to only a small sub-sample of the globular cluster
system and is therefore subject to various bias effects, as will be
discussed later. In this respect we also note the most recently
published results on NGC~4365 by Brodie et al (2005). In this paper
spectral data of GCs are used to derive their ages, which seem to
contradict previously obtained age estimates, both spectroscopic and
photometric, and claim an average age of 11 Gyr for the complete GC
sample, with no evidence for intermediate age clusters. At this point
we have to take into account that the different samples have only few
globular clusters in common. However, the authors discuss the
possibility of spatial distribution effects, which is indeed an
important issue when drawing conlusions on ages of selected
samples. Clearly more data are required to explain the different
results.

\begin{figure*}[!ht]
\center
\includegraphics[width=8cm]{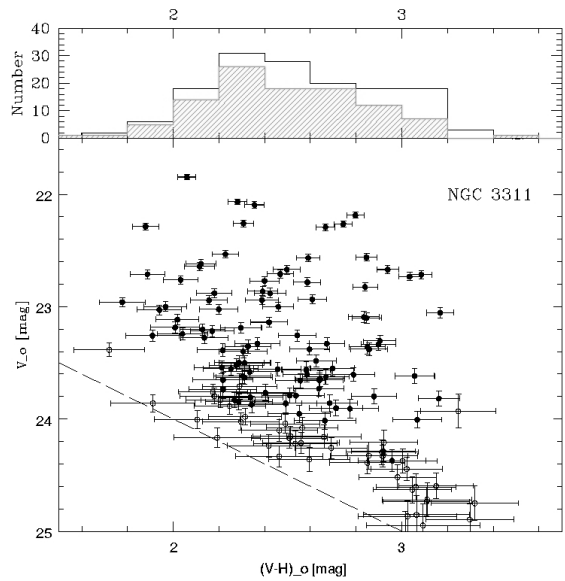}
\caption[$V~vs.(V-H)$ color-magnitude diagram for
NGC~3311]{$V~vs.(V-H)$ color-magnitude diagram for NGC~3311. The top
sub-panel shows the color distribution of all (open histogram) and
error-selected (hashed histogram) objects. The lower sub-panel shows
the color-magnitude diagram. Herein the filled symbols mark the
selected clusters, while the open circles mark rejected objects (by
error cut).  The dashed line marks the 50$\%$ completeness limit for
the H-band, derived from the detection rate in artificial data sets
using the Addstar routine in the IRAF package.}
\label{n3311f3nic}
\end{figure*}

\begin{figure*}[!ht]
\center
\includegraphics[width=8cm]{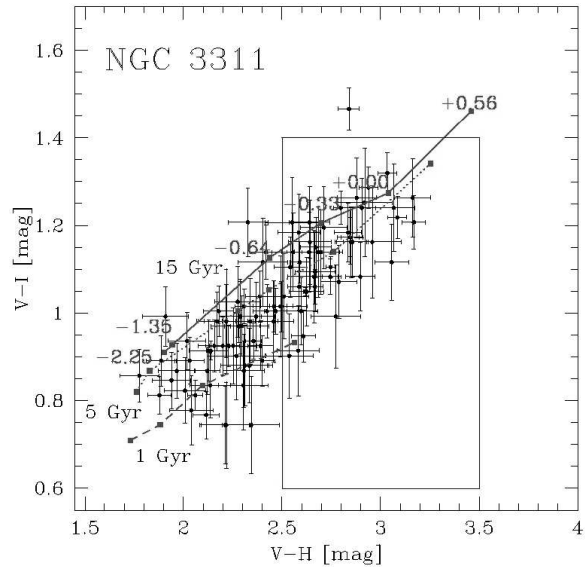}
\caption[$(V-I)$~$vs.$~$(V-H)$ color-color diagram for
NGC~3311]{$(V-I)$~$vs.$~$(V-H)$ color-color diagram for NGC~3311. The
data obey the selection by photometric error cut ($<$0.15~mag) and are
corrected for foreground reddening (see Table~\ref{parametersnic}).
As examples the 15 Gyr, 5 Gyr, and 1 Gyr isochrones, corresponding to
the Bruzual \& Charlot model isochrones (\cite{bruz03nic}) are
marked. The metallicity, rising from [Fe/H]=-2.25 to [Fe/H]=+0.56, is
marked by solid squares. The color range selected for the
determination of the cumulative age distribution is marked by the
box. Compared to IC~4051 the $(V-H)$ color range is slightly shorter
to fit the observed data.}
\label{n3311colcolnic}
\end{figure*}

At this point we want to emphasize that only within the red
sub-population is it possible to distinguish the age of globular
clusters whereas the blue population should represent only the first
generation, but deriving even crude age estimates for such blue clusters is
very difficult. This supposedly old fraction is NOT included in
the derived size of the different age populations. In Section
\ref{introductionnic} we mentioned a gas rich merger as a possible
scenario to form a second generation of globular clusters, which
requires a certain minimum galaxy density. IC~4051 is found somewhat
outside the core of the Coma galaxy cluster, surrounded by numerous
other galaxies, but not in the center of the gravitational potential
of the galaxy cluster. However, additional evidence for a less passive
evolution of this galaxy is the existence of a co-rotating galaxy
core, which rotates at a different velocity from the main galaxy
body. It is not known yet whether it is dynamically decoupled or not
(\cite{mehlert98nic}). Such peculiarities are quite common in
early-type galaxies (e.g. \cite{davies01nic}, \cite{zeeuw02nic}) and
we do not intend to link the existence of such a co-rotating core
directly with a second star formation event. However, if the dynamical
structure of an early-type galaxy sets it apart from the formerly-
uniform and simple structured picture, then these will be prime
targets to look for multimodal age distributions. In this respect we
want to emphasize that our classification of IC~4051 as an early-type
galaxy in a less dense environment is rather relative, since the Coma
galaxy cluster represents a high density environment (projected central galaxy
density of $\approx$39 galaxies/Mpc$^2$,
\cite{kent82nic}) compared to the Hydra I galaxy cluster (17
galaxies/Mpc$^-2$, \cite{richter89nic}). Nevertheless, due to its
distance of 15\farcm5 ($\sim$460~kpc, see \cite{jorgensen94nic}) from
the cluster center we do not consider IC~4051 as a galaxy in a high
density environment anymore.\\

\vskip 0.2cm
\subsubsection{NGC~3311}
\label{resultn3311nic}
The $V~vs.~(V-H)$ color-magnitude diagram and $(V-I)$~$vs.~$$(V-H)$
color-color diagram for NGC~3311 are presented in Figures
\ref{n3311f3nic} and \ref{n3311colcolnic}. The histogram in
the upper part of the CMD shows the color distribution of the complete
sample as an open diagram, whereas the error selected sample is shown
by the hatched histogram.

The color-color distribution in Figure \ref{n3311colcolnic} shows
that in addition to many old GCs, a large fraction of red globular
clusters ($V-H~> 2.5)$ have a $(V-I)$ color which is too blue to be
consistent with an old cluster population. A direct comparison to the
SSP isochrones assigns those objects an age $\lesssim$5 Gyr. The
possibility of a young/intermediate population of globular clusters
formed in a merger was already discussed by Hilker (2002). Our
near-infrared observations do not favor the second possibility,
viz. an old and metal poor population. Interestingly, NGC~3311
contains a prominent dust lane in its center
(e.g. \cite{vasterberg91nic}; \cite{grillmair94nic}), which adds
further evidence to support a recent merging or accretion event. The
dust content in the core of NGC~3311 was estimated as
10$^{3.5}$~M$_{\odot}$ (\cite{ferrari99nic}) based on V and R-band
photometry. Grillmair et al. (1994) derived a lower limit of
4.6$\times$10$^{4}$ M$_{\odot}$ for the total dust mass. Assuming a
similar gas-to-dust ratio as found in the Galaxy,
{\it{m$_{gas}$/m$_{dust}$}}$\sim$100 (\cite{jura86nic};
\cite{mcnamara90nic}), the total gas and dust mass will reach at least
4.6$\times$10$^6$M$_{\odot}$. The globular cluster sample included in
our analysis, however, refers only to the innermost region in
NGC~3311. Additional near-infrared observations covering larger
galactocentric radius will be necessary before any conclusion can be
inferred from comparing our small central sample to the entire blue
globular cluster population found by Hilker (2002), using optical
photometry alone. As previous
studies on NGC~4365 (\cite{puzia02nic}) have shown, results based on
optical photometry alone do not always indicate the existence of age
sub-populations. The distribution of $(V-I)$ colors shows only the
common bi-modal distribution, mostly caused by the metallicity. Only
the combination with near-infrared colors reveals the more complex age
structure of the globular cluster system.\\

\vskip 0.5cm
\subsection{Cumulative Age Distribution}
\label{agenic}

\begin{figure*}[!ht]
\center
\includegraphics[width=8cm]{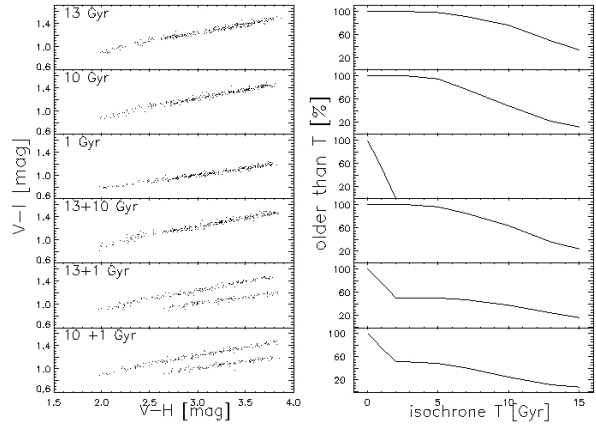}
\caption[Simulation of color-color diagrams and the resulting age
distribution]{Simulated color-color diagrams (left) and the resulting
cumulative age distribution (right) for various age combinations. The
two populations are equally mixed between the old and intermediate age
populations. The ideal colors, following the Bruzual \& Charlot (2003)
SSP models, were smeared randomly with an up to 3$\sigma$ photometric
error.}
\label{modelnic}
\end{figure*}

Our technique to derive cumulative age distributions, as described in
Hempel et al. (2003) and Hempel \& Kissler-Patig (2004), is to help
lift the age-metallicity degeneracy (\cite{worthey94nic}) via the
combination of optical and near-infrared broad band colors. Hereby the
higher sensitivity of infrared colors to metallicity as compared to
age is utilized. An example of the comparison of the age and
metallicity dependency of $(V-I)$ and $(V-K)$ ($\sim (V-H)$) can be
found in Puzia et al. (2002) and references therein. The cumulative
age distribution is then defined as the percentage of objects
{\bf{older than T Gyr}}, and is derived by comparing their $(V-I)$
color with the SSP model prediction for a population that is {\bf{T
Gyr}} old. The lowest age bin is assigned to 0 Gyr and includes all
GCs within the selected $(V-H)$ color range. With respect to a
possible formation scenario of the GCSs and their host galaxies, we
want to go one step further and set constraints on the age and the
relative size of the cluster sub-populations. The method is based on
the comparison between observed and simulated GCSs
(\cite{hempel03nic}; \cite{hempel04anic}) via a $\chi^2$ -test. To
simulate the color distribution in GCSs and to derive the age
distribution in simulated and observed systems, the SSP model
isochrones of Bruzual \& Charlot (2003) were again used. Examples for
the simulated color-color diagrams and the resulting cumulative age
distributions are shown in Figure \ref{modelnic}. Instead of a
step-function-like age distribution we find a much more gradual slope
in the age distribution, which is mostly due to the fact that the
$(V-I)$ colors in our simulations are not strictly independent from
metallicity effects and, therefore, show a certain dependency on
$(V-H)$. The final consequence is that even with a much improved age
resolution of combined optical and near-infrared colors, the
determination of accurate absolute ages for single globular clusters
is still out of reach. Our intention is to detect age sub-populations,
to set constraints on their age and to find systematic effects of
external parameters on the age structure. In nearby galaxies more
precise age determinations can be derived from spectroscopic spot
checks. For our simulations we assume a composition of two age
sub-populations with ages ranging from 1, 1.5, 2, 3, 5, 7, 10 and 13
Gyr. Hereby we vary the amount of young/intermediate age globular
clusters between 100$\%$ and 0$\%$. The $\chi^2$- test finds the best
fitting model with a given age and size ratio.\\

\begin{figure*}[!ht]
\center
\includegraphics[width=8cm]{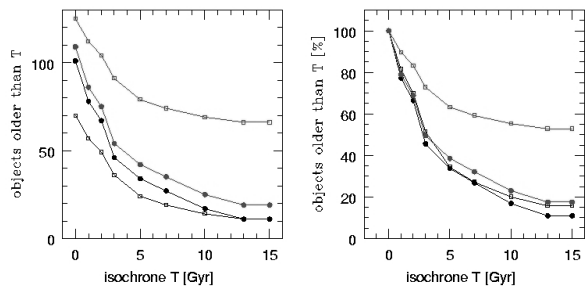}
\caption[Cumulative age distribution in NGC~3311 and
IC~4051]{Cumulative age distribution in NGC~3311 (open
squares) and IC~4051 (solid circles) derived following the procedure
given in Section \ref{agenic}. The left panel shows the absolute
numbers of globular clusters older than a given age. In the right
panel the distribution was normalised with respect to the total number
of objects found. If we assume the blue population of globular
clusters ($(V-H)$$<2.5$) to consist of only 13 Gyr old objects the
cumulative age distribution would change as shown by the grey
patterns.}
\label{agedisgalnic}
\end{figure*}

As shown in Sect. \ref{resultn3311nic} and \ref{resultic4051nic} the
color-color diagrams give us a first indication for the existence of
multiple age sub-populations in the GCSs of both NGC~3311 and
IC~4051. If we compare the cumulative age distributions (Figure
\ref{agedisgalnic}) with the simulation of a purely old system (see
Figure \ref{modelnic}, upper, right panel) the differences are
obvious.

From Figure \ref{agedisgalnic}, the age distribution in both GC
samples assigns $\sim$65$\%$ of the globular clusters to an age
younger than 5 Gyr. Recall that this refers only to GCs with
$V-H>2.5$. Although the combination of optical and near-infrared
colors helps lift the age-metallicity degeneracy, $(V-I)$ is still
mildly affected by the metallicity. Combined with the scatter in both
colors, induced by the photometric errors, and the relatively small
sample size (IC~4051: 101 GCs, NGC~3311: 70 GCs), this results in a
significant uncertainty with respect to the size estimate of the
sub-populations. To overcome this obstacle we will not use the
cumulative age distribution as a direct measure for the age structure
in the GCSs, but rather compare the observations with the simulated
systems via a $\chi$$^2$- test. In this test we determine the
difference between the cumulative age distribution in observed and
simulated systems. Nevertheless, we still have to accept a certain
degeneracy between age and size of sub-populations, which results in a
finite accuracy with respect to the best fitting model. Since the
major goal of this project is to detect age sub-populations and set
some constraints on their age, it is of no importance here (and
impossible for us to detect) whether the second generation of globular
clusters is 2 or 3 Gyr old and contains 30 or 40 $\%$ of the globular
cluster sample. This uncertainty becomes even more important for NGC~3311, where a large
fraction of the observed GCs are excluded from the age analysis by the
color cut. In the comparison of our results with previously obtained
age estimates we will therefore concentrate on IC~4051
(e.g. \cite{mehlert98nic}, \cite{mehlert03nic}). Based on
spectroscopic data (Figure 8 in \cite{mehlert98nic}) the age of the
galaxy core of IC~4051 was determined as 12-17 Gyr, based on the
Worthey SSP models (\cite{worthey94nic}). We note that agreement with
model predictions was only obtained for the innermost 5\farcs0 of the
galaxy core, for which the photometric data of the globular clusters
suffer from contamination by the diffuse galaxy light. An additional
error source is the fact that in the diffuse galaxy light the
contributions of stellar sub-populations are blended and hard to
disentangle.  \\

\begin{figure*}[!ht]
\center
\includegraphics[width=7cm]{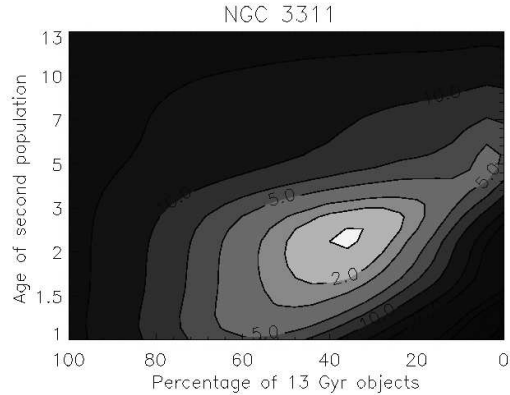}
\caption[$\chi^2$- test result for NGC~3311]{$\chi^2$- test result for
NGC~3311. Different levels represent the reduced $\chi^2$ of the
comparison between the cumulative age distribution in NGC~3311 and
various models. The best fitting model consists of $\sim60\%$
intermediate age globular clusters and $\sim$40$\%$ of 13 Gyr old
objects. The large size of this young population is surprising and is
discussed in Section \ref{s:summarynic}.}
\label{n3311contournic}
\end{figure*}

\begin{figure*}[!ht]
\center
\includegraphics[width=7cm]{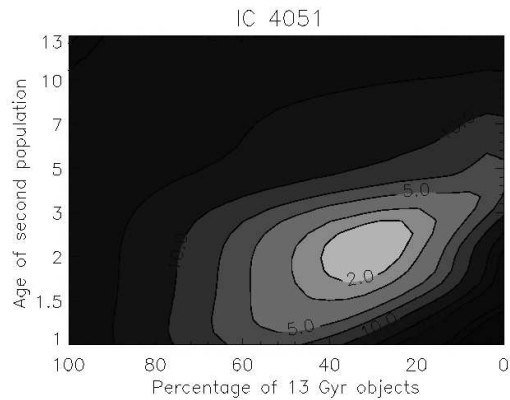}
\caption[$\chi$$^2$-test for IC~4051]{The comparison between the age distribution in IC~4051 and
the simulated systems again (as in Figure \ref{n3311contournic})
reveals a large fraction of young/intermediate age objects
($>$$\thicksim$60$\%$), with the population of second generation globular
clusters slightly larger than in NGC~3311.}
\label{ic4051contournic}
\end{figure*}

The results of these $\chi$$^2$ tests are presented in Figures
\ref{n3311contournic} and \ref{ic4051contournic}. The $\chi$$^2$ test
results find the best fitting model to the age distribution in both
GCSs to be a composite of an old population and a second population
which is $\sim$10 Gyr younger. In both galaxies, the best fitting
model to the age distribution suggests approximately 60$\%$ to 70$\%$
intermediate age clusters. Such a surprisingly large fraction of
young/intermediate globular clusters, also found in NGC~4365 and
NGC~5846 (\cite{hempel04anic}), needs confirmation and careful
investigation, especially since such large fractions of young stars
should be visible in the integrated light, but are not detected
yet. In fact, previous investigations based on optical photometry
(\cite{woodworth00nic}), suggest the bulk of the GCS being formed in
situ, which, in the authors understanding, also includes the
possibility of merging of almost completely gaseous objects. A late
gaseous merger is not in any way ruled out using as well the argument
of the central co-rotating disk and its very high metallicity.\\

The major goal in this work is to apply a semi-numerical method for
detecting globular cluster sub-populations based on optical and
near-infrared colors. As described in Hempel et al. (2003) and Hempel
\& Kissler-Patig (2004) this procedure has been developed in order to
partly compensate for the photometric uncertainties, increased by the
usage of near-infrared observations, when deriving age
constraints. Instead of a direct comparison between the colors of each
individual observed globular cluster with model predictions, as
attempted with color-color diagrams (see Figures \ref{ic4051f6nic} and
\ref{n3311f3nic}) we rather look at the density of GCs with respect to
a set of model isochrones and compare it to simulated GCSs. Although
the results of both approaches may differ in the actual age estimates,
the existence of distinct age sub-populations should be revealed by
both approaches, the semi-numerical one being less sensitive to
photometric errors. Again, concentrating on the IC~4051 results shown
in the color-color diagram (see Figure \ref{ic4051f6nic}) we are
unable to derive individual globular cluster ages by comparing the GC
colors directly with the SSP model predictions, but nevertheless find
a large fraction of globular clusters with $(V-I)$ colors being better
represented by a isochrone for a 5 Gyr population or even younger. A
better agreement between the age estimate via our semi-numerical
approach should be possible by refining the age grid for the
Monte-Carlo simulations (e.g. \cite{hempel04anic}) as well as for the
determination of the cumulative age distribution.

The derived age structure will therefore be affected by different
factors, the most important being stated below:
\begin{itemize}
\item{\bf{Firstly}}, we have to consider that the sample used in our
analysis is color-selected, so only objects with $2.5\leq~(V-H)\leq~3.5$
and $0.6\leq~(V-I)\leq~1.4$ are used. The lower limit to $(V-H)$ is
set due to the small color differences in the SSP models. The blue,
metal-poor population of globular clusters will mostly contribute to
the old sub-population and change the size ratio between old and young
population in favor of the first generation objects. The upper
limit, as well as the $(V-I)$ interval, is set to avoid significant
contamination of our sample with background galaxies
(\cite{puzia02nic}). Especially the $(V-H)$ blue fraction, supposedly old
globular clusters, will affect this result and shift the ratio between
old and young/intermediate age globular clusters to higher
values. However, as shown in Figure \ref{agedisgalnic}, even in the
extreme case where all GCs bluer than $(V-H)$=2.5 are old, 13 Gyr old
clusters there remains a substantial fraction of
young/intermediate age clusters.\\
\item{\bf{Secondly}}, we have to be aware of spatial bias effects. Due
to the small NICMOS field of view, centered on the galaxy, the data
cover only the innermost region of NGC~3311 and IC~4051. As observed
by Schweizer \& Seitzer (1998) and simulated by Hibbard \& Mihos
(1995) for the merger remnant NGC~7252, the central region of a galaxy
is where one would expect to preferentially find the second generation
of globular clusters, if they are indeed formed during a merger or
accretion event. Although mostly born in the tidal tails
(\cite{li04nic}) of a merger, these star clusters will funnel toward
the center of the merger remnant within several 100 Myr up to 1 Gyr
(\cite{hibbard95nic}). Our samples are therefore spatially biased and
the fraction of young/intermediate age globulars should be
overestimated, compared to the global fraction over the entire
galaxy.\\
\item{\bf{Thirdly}}, small number statistics will affect the $\chi^2$-
test. As shown in Hempel \& Kissler-Patig (2004) the results of this
test become unstable (i.e. show multiple best fitting age/size
combinations) for small data sets. Previous tests have shown that
$\gtrsim$100 objects are needed to obtain reliable results. Strictly
speaking this condition is only (barely) achieved in the IC~4051
sample. Recall that the total sample of IC~4051 and NGC~3311 contain
256 and 148 objects, respectively, but after applying the selection
criteria (photometric error and color cuts) only 101 (IC~4051) and 70
(NGC~3311) objects contribute to the age distributions. Although the
presence of a significant population of young/intermediate age
globular clusters in NGC~3311 agrees with various other features of
the galaxy (e.g. central dust lane) more data are required for a solid
argument.  
\item{\bf{Fourthly}}, as mentioned in Section \ref{introductionnic}
the SSP model isochrones by Bruzual \& Charlot (2003) have been
applied. As shown by Hempel et al.(2003) and Hempel \& Kissler-Patig
(2004) the color predictions for a given age by various SSP models
differ particularly in the red color range and model dependencies
should be considered and we will do so in detail in Appendix
\ref{appendixa}. Since our method is based on the object density in
color-color diagrams with respect to a given set of SSP isochrones,
the final result, i.e. the detection of age sub-populations will hold
independently of the SSP model. The number of objects required for a
stable result of the $\chi$$^2$-test is however strognly dependent on
the $\delta$age/$\delta$$(V-I)$ gradients. The models by Maraston
(2005) are somewhat hampered by the fact that stellar populations with
ages below 5 Gyr (below 2 Gyr for the 2005 release) show a
non-monotonic increase of $(V-I)$ with $(V-H)$ relation, which leads
to confusion in our counting algorithm. The major advance in the more
recent SSP models are the inclusion of a well developed AGB phase
(e.g. Renzini~1981, Maraston et al.2001b), the possibility of a
different [$\alpha$/Fe]-enhancement (\cite{thomas03bnic}) and the
structure of the horizontal branch (\cite{maraston05nic}), resulting
in color shifts. Although $\alpha$- enhancement seems to be more
important when objects with solar metallicity are investigated no
solid conclusions can be drawn, due to less solid calibration of the
models. We will include the latest release of the Maraston models
(2005), targeting the influence of the horizontal branch structure in
an additional section (see Appendix \ref{appendixa}).
\end{itemize}

\section{Metallicity Distribution}
\label{s:metalnic}

In the framework of galaxy formation and evolution studies the age
structure of the galaxy and its GCS plays a major role hence our main
interest is focused on the age structure. Nevertheless, since the
latter is determined by using SSP model isochrones, which combine age
and metallicity predictions, it is necessary to test to what degree
age and metallicity effects are still entangled. In the following we
present the results of rough metallicity estimates based on the
derived age structure.

A 2 Gyr old stellar population, with a $(V-H)$ color between
$2.5\leq(V-H)\leq3.5$, i.e. consistent with the color limits set for
the determination of the age distribution, contains objects with
metallicities between [Fe/H]=+0.00 and [Fe/H]=+1.1, respectively, as
derived via linear interpolation using the SSP models. Assuming a mean
color of $(V-H)$=3.0 we obtain [Fe/H]=+0.56. Although, as shown in
Figure \ref{ic4051contournic}, the best fitting model consists of a 2
Gyr old second globular cluster population, the age uncertainty allows
as well a higher age, i.e. $\sim$5 Gyr. In this case the metallicity
of objects in the same color range lies between [Fe/H]=-0.2 and +0.9,
with a mean value of [Fe/H]=+0.4 for $(V-H)$=3.0. As we will show,
this corresponds closely to the mean metallicities determined for the
young/intermediate GCs in IC~4051. Since the metallicity depends quite
strongly on the derived age (more than the other way around) and
although we can detect age sub-populations and set constraints on the
age, the age uncertainty allows a large variety of metallicities. The
following discussion about the metallicity distribution has to be seen
in the light of this persistent degeneracy between age and
metallicity.

With the introduction of combined optical and near-infrared
observations we are able to investigate the metallicity distribution
approximately independently from the age structure, with the above
caveats. In order to derive globular cluster metallicities we use the
above-derived information about the age structure in the observed
globular cluster sample (see Figures \ref{n3311contournic} and
\ref{ic4051contournic}). The results in Section \ref{agenic} tell us that
the age structure in both systems is well matched by a mix of a 13 Gyr
old population with globular clusters which are $\sim$10 Gyr
younger. The determination of the metallicity distribution follows the
procedure given below and applies the Bruzual \& Charlot (2003) SSP
model isochrones.

\begin{figure*}[!ht]
\center
\includegraphics[width=12cm]{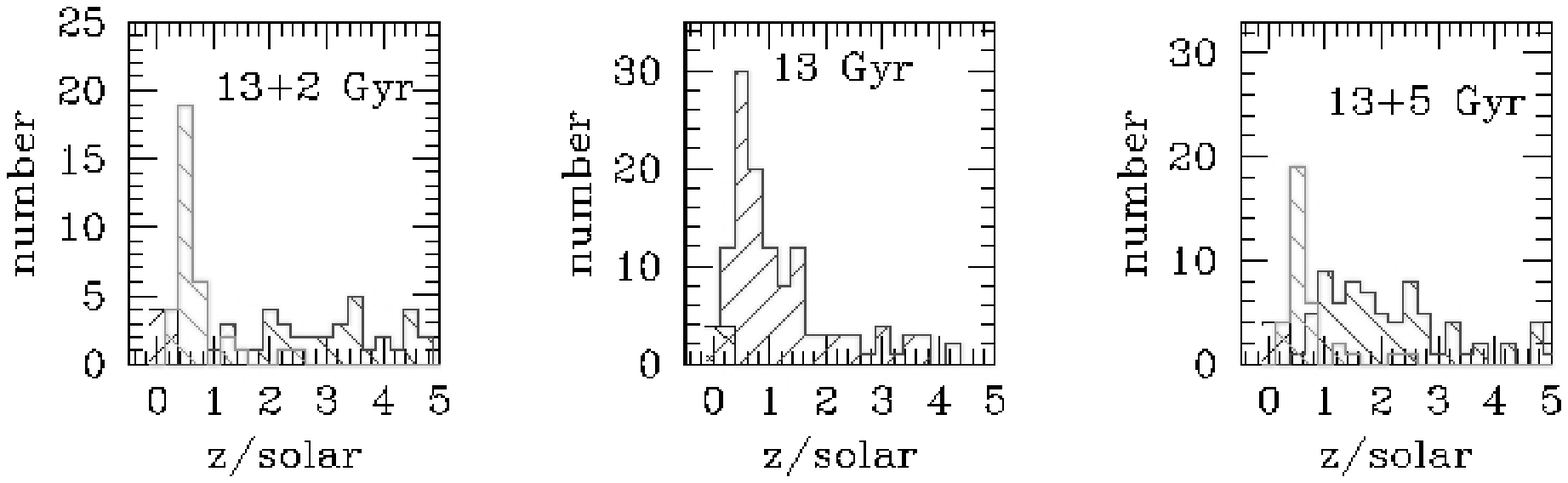}
\includegraphics[width=12cm]{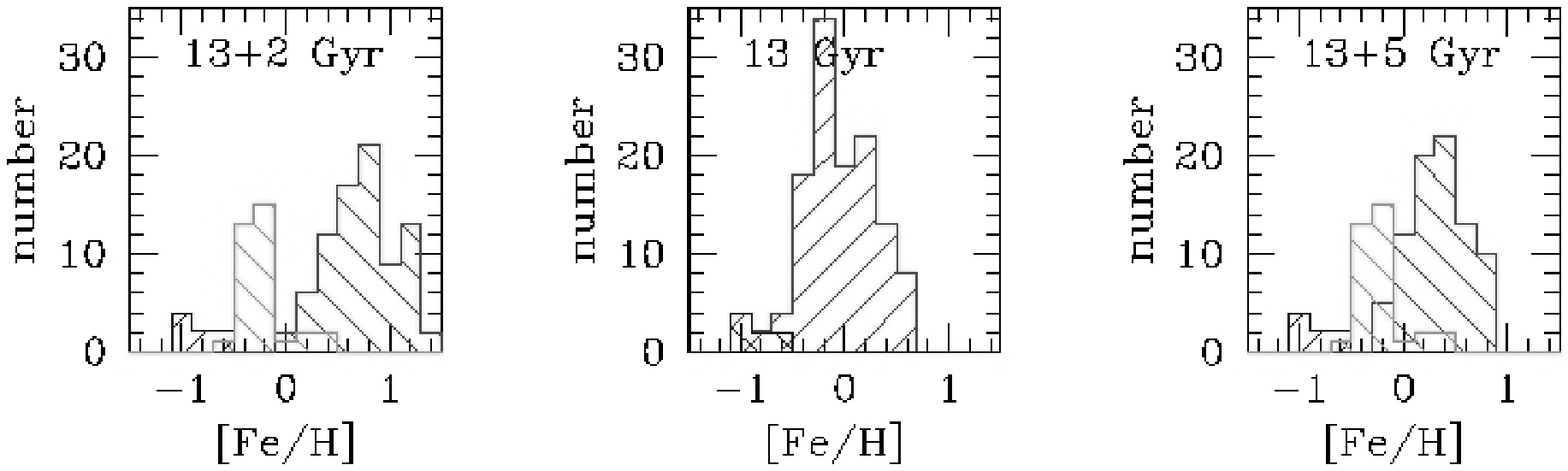}
\caption[Metallicity distribution in IC~4051]{Metallicity distribution
in IC~4051 assuming different GC distributions. We show the
metallicity distribution both in terms of linear metal abundance
(upper panel) as well as in logarithmic metallicity [Fe/H] (lower
panels). In the left panels the metallicity distribution is based on
the assumption of a 2 and 13 Gyr old population (following the best
fit model from the age distribution), whereas the central panel
assumes all clusters to be 13 Gyr old. We also show the metallicity
distribution assuming the young/intermediate globular clusters to be 5
Gyr old (right panel). The color coding of the histograms is as
follows; dark grey: young and metal rich, light grey: old and metal rich, black:
old and metal-poor.}
\label{metal_ic4051nic}
\end{figure*}

\begin{figure*}[!ht]
\center
\includegraphics[width=12cm]{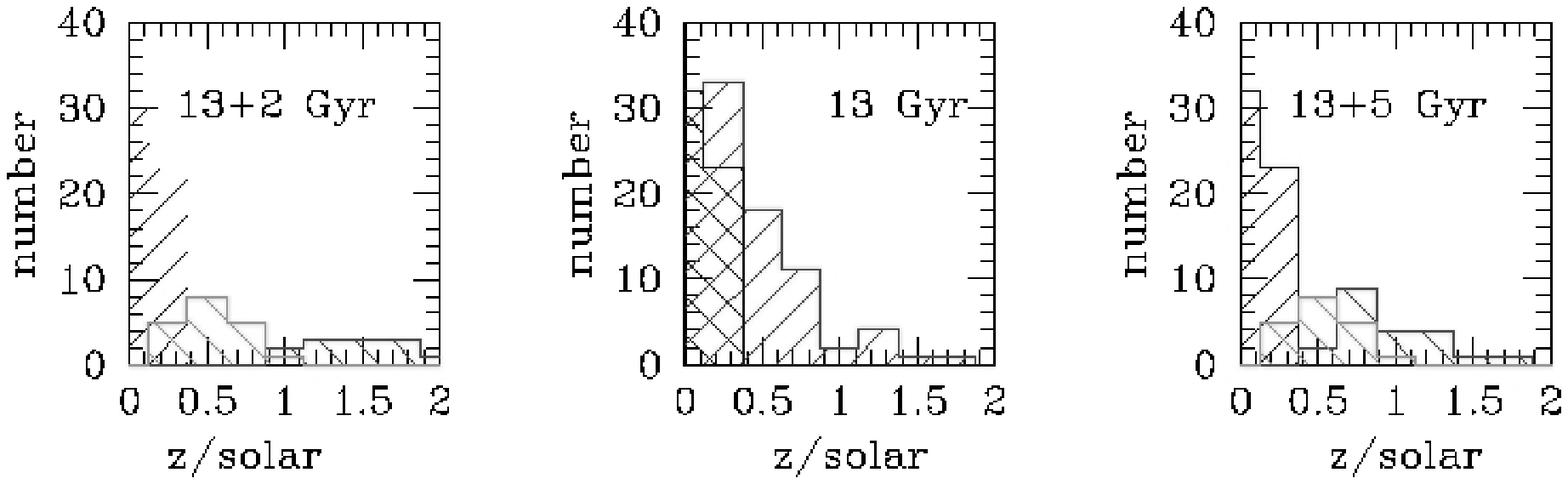}
\includegraphics[width=12cm]{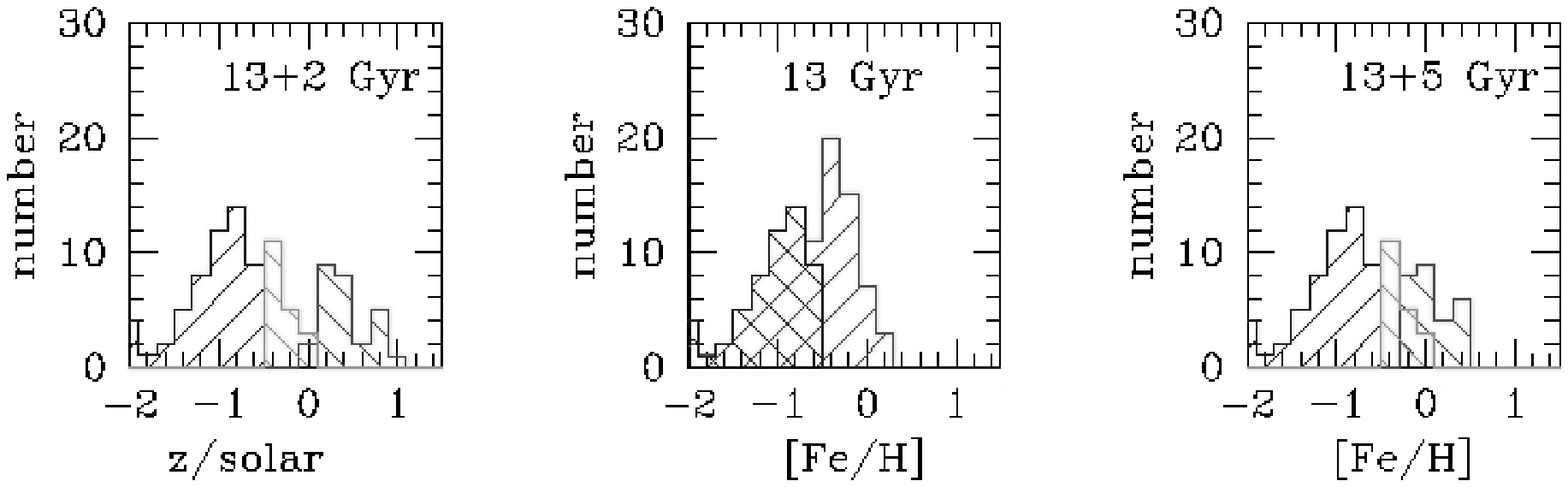}
\caption[Metallicity distribution in NGC~3311]{Metallicity
distribution in NGC~3311 assuming different age distributions. We
apply the same color coding and method for displaying the abundance
distribution as in Figure
\ref{metal_ic4051nic}.}
\label{metal_n3311nic}
\end{figure*}

\begin{itemize}
\item We split the globular cluster sample (selected by photometric
error) with respect to the $(V-H)$ color into a blue ($(V-H)$$<2.5$)
and presumably old population and a second, red ($2.5<(V-H)$)
sub-population, containing both old as well as young/intermediate globular
clusters. In addition we set an upper limit of $(V-H)$$\le3.5$ to
avoid strong contamination with background galaxies
(\cite{hempel04anic}).
\item For globular clusters found in the blue sub-population, the
metallicities (given as Z/Z$_{\odot}$) were calculated by
interpolating the metallicity-$(V-H)$ correlation of the 13 Gyr model
isochrone.
\item Corresponding to the age structure, we found that the red sub-population is
split into old and intermediate/young populations, and use the 5 Gyr
model isochrone for the latter. All objects with
$(V-I)$~$\le$~$(V-I)_{5~Gyr}$ are then assumed to form the second
generation of globular clusters. Although the comparison of the
cumulative age distributions to simulated GCSs assigns the
intermediate/young objects to an age of $\sim$2 Gyr, we allow an age
uncertainty of 3 Gyr, based on the $\chi$$^2$ contours, and set the
age split at 5 Gyr.
\item The metallicity is then determined by interpolating
Z/Z$_{\odot}$ at $(V-H)_{GC}$ using the 13 Gyr and 2 Gyr isochrones (Figures
\ref{metal_ic4051nic} and \ref{metal_n3311nic}, left panels). For 
comparison we also derive the metallicity distributions assuming a
purely old globular cluster population (Figures \ref{metal_ic4051nic}
and \ref{metal_n3311nic}, middle panels) and for mixed GCSs containing
13 Gyr and 5 Gyr old objects (Figures \ref{metal_ic4051nic} and
\ref{metal_n3311nic}, right panels). In both Figures the upper panels
give the metallicities on a linear scale, whereas the lower panels
refer to [Fe/H] values.
\item The metallicity estimates depend strongly on the age structure and are
therefore also hampered by bias effects (see Section
\ref{agenic}). The case of NGC~3311 suffers from small-number
statistics, which makes the results of the $\chi$$^2$-test
statistically unstable. Nevertheless, the comparison of the $(V-H)$
colors, seen in Figures \ref{ic4051f6nic} and \ref{n3311colcolnic},
shows that NGC~3311 lacks the most metal rich globular clusters found
in IC~4051. Due to the above uncertainty for NGC~3311 we will
concentrate on the GCS in IC~4051.
\end{itemize}

\begin{figure*}[!ht]
\center
\includegraphics[width=12cm]{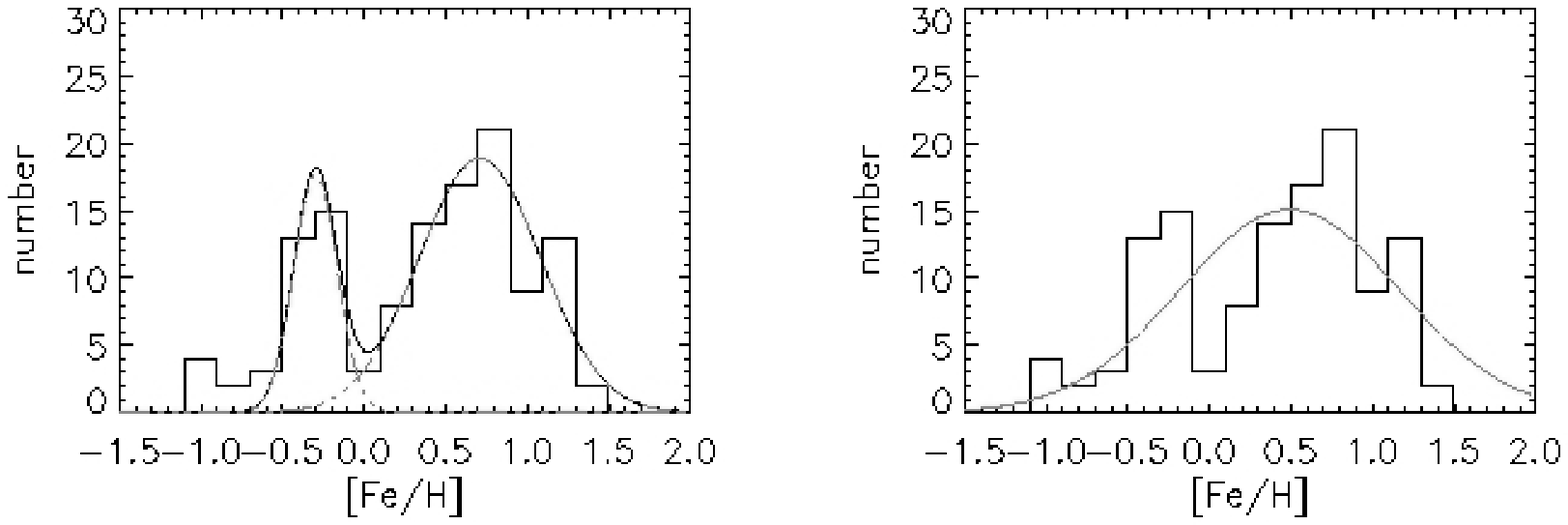}
\includegraphics[width=12cm]{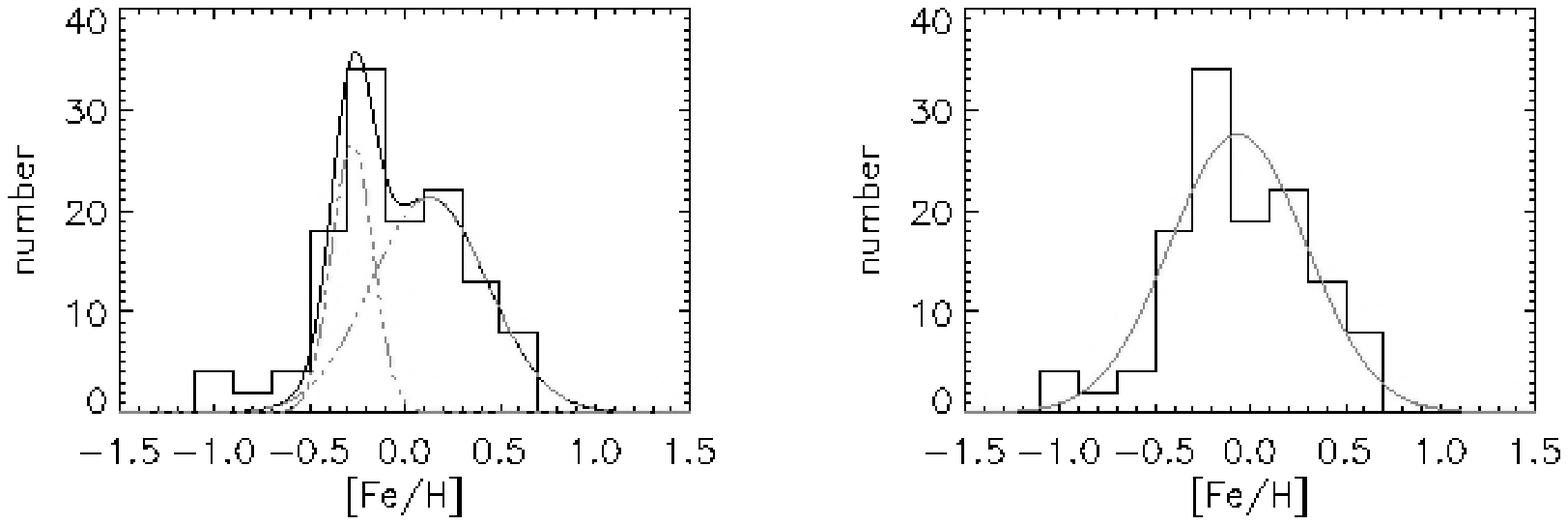}
\includegraphics[width=12cm]{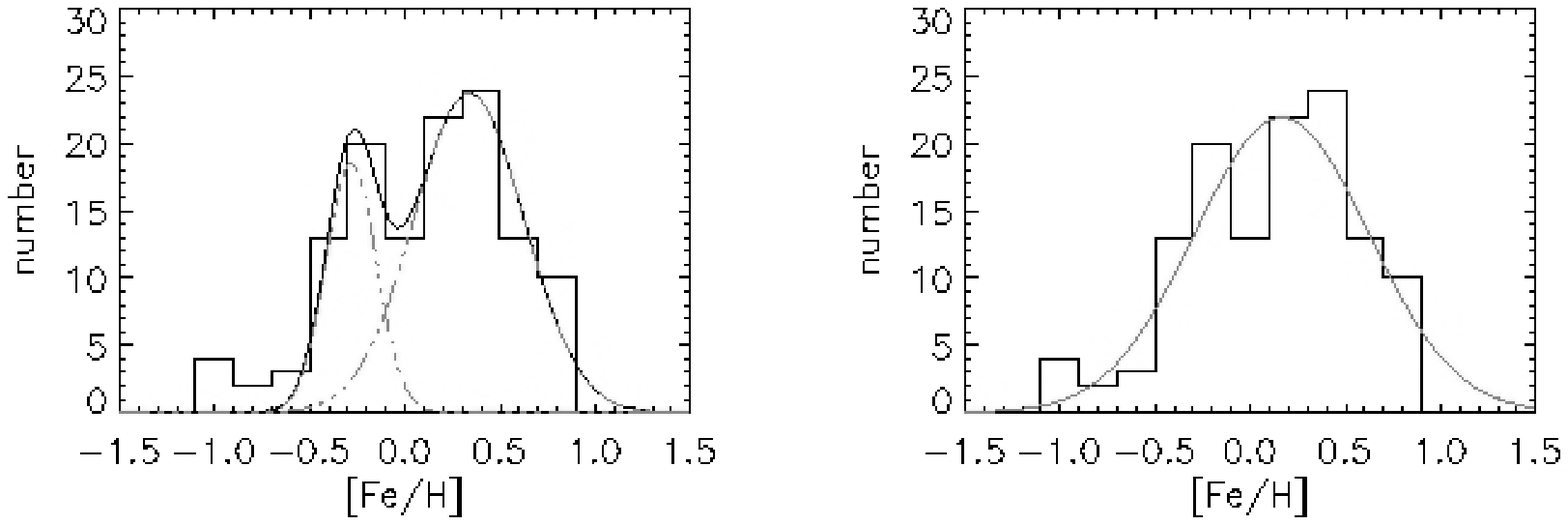}
\caption[Gaussian fit to metallicity distribution]{Metallicity
distribution in IC~4051 assuming a combination of a 13 and 2 Gyr old
population (upper panels), a pure 13 Gyr old population (middle
panels) and a 13 +5 Gyr old mixed population (lower panels). On the
left we show the best bimodal Gaussian fit while on the right that for
a single Gaussian. Clearly the mixed populations are best fitted by a
double Gaussian (dashed-dotted lines). The fit to the metallicity
distribution of a 13 Gyr population gives similarly good results for a
single and a double Gaussian, although the $\chi$$^2$ of the fit is
slightly better for a single Gaussian. In the left panel the
dashed-dotted lines show the two Gaussians whereas the solid line
shows their combination.}
\label{metal_ic4051fit}
\end{figure*}

In the discussion about the bimodality of metallicity distributions we
want to emphasize the importance of the chosen scaling. It is
interesting to note that the metallicity distribution for a purely old
(13 Gyr) population seems to suggest a bimodality in the [Fe/H]
distribution for the IC~4051 GC, which is NOT seen if we transform the
GC metallicities into linear abundances. However, if we fit the [Fe/H]
distribution by either a double or a single Gaussian, as shown in the
middle panels in Figure \ref{metal_ic4051fit}, the resulting $\chi^2$
of the fit is found to be $\chi^2$=0.6 and 0.8 for the double and
single Gaussian, respectively, virtually identical. Hence there is no
strong evidence for bimodality in the metallicity distribution, which
considers all clusters as old. Although a bimodal distribution is not
favored over unimodality in the 13 Gyr old population, it becomes
obvious in the samples assuming mixed globular cluster age
distributions (see upper and lower panel in Figure
\ref{metal_ic4051fit}).

Since we find objects within a wide range of metallicities, applying a
logarithmic scaling, as given in the distribution of the [Fe/H]
values, increases the chances for detecting metallicity
populations. In the linear Z space formerly clearly detected
[Fe/H]-populations are much more spread out and may not be detected
anymore, e.g. in smaller samples or in globular cluster populations
with a smaller age difference. The metallicity distribution given in
Z/Z$_{\odot}$ for a 13 + 2 or 5 Gyr old population is widely spread.  We
caution that a bimodal color distribution does not necessarily
translate into a clear bimodal metallicity or linear abundance
distribution (see also
\cite{kisslerpatig02nic}).

The mean metallicity based on the $(V-I)$ colors alone and a unimodal
color distribution (\cite{woodworth00nic}) was determined to be
$\langle$Fe/H$\rangle$=-0.3 or Z/Z$_{\odot}$=0.5. Working with the
more metal sensitive $(V-H)$ color we obtain the following
results. Assuming a purely 13 Gyr old population in IC~4051 (see
Figure \ref{metal_ic4051nic}, central panel), we derive a mean
metallicity of $\langle$Fe/H$\rangle$$\sim$-0.1, assuming a unimodal
distribution of the metallicity. For metal-poor AND a metal
rich-population of the same age we obtain mean metallicities of -0.27
and +0.12 dex, respectively. For a mixed-age GCS, the mean metallicity
of the 2 Gyr old population is derived as $\langle$Fe/H$\rangle$$_{2
Gyr}$=+0.7, with $\langle$Fe/H$\rangle$$_{13 Gyr}$=-0.9 and -0.3 for
the metal-poor and metal-rich old population. Assuming 5 Gyr as the
age of the intermediate age population, the results change only
marginally to $\langle$Fe/H$\rangle$$_{5 Gyr}$=+0.34. The
metallicities derived for the young metal-rich populations in either
case are very high; compared to observed spectroscopic metallicities
for other second generation GCs (e.g. \cite{schweizer98nic},
\cite{goudfrooij01anic}, \cite{goudfrooij01bnic}) and must be regarded
as very tentative. Several issues have to be discussed here, which all
affect the derived metallicity distribution.
\begin{itemize}
\item Again, the basic requisite to derive globular cluster metallicities is
an accurate age determination. Although the combined optical and
near-infrared photometry allows us to derive the mean age of
sub-populations, the age-metallicity degeneracy is not entirely
broken. Even more important, for a given $(V-H)$-bin the derived
metallicity depends on the SSP isochrone. Thus applying the mean age
of sup-populations to calculate individual metallicities causes
additional systematic errors.
\item In our analysis we use the SSP model isochrones by Bruzual \&
Charlot (\cite{bruz03nic}). If we compare the
{\bf{color}} predictions of these models with Vazdekis (1999) or
Maraston (2001) then we find considerable differences, especially
within the red populations. Following the Bruzual \& Charlot models a
5 Gyr old object would by $\sim$0.1 mag redder in $(V-I)$ compared to
the Vazdekis models, given the same $(V-H)$. This discrepancy has only
little effect on the mean ages of globular cluster sub-populations as
long as we apply identical SSP models for the age determination in
observed and simulated samples and do not attempt to derive ages of
individual clusters. Therefore, our primary result of a significant
number of intermediate age globular clusters will not change. The fact
that its age alters with the applied SSP models between 2 and 5 Gyrs
is of no relevance since the method itself does not allow more precise
age estimates. To quantify the effect of our choice of SSP models on the
metallicity distribution we will repeat the analysis, i.e.  the
determination of the cumulative age distribution and the modeling
procedure, using the SSP model isochrones by Maraston (2005) (see Appendix \ref{appendixa}).
\item In Paper IV of this series (Hempel \& Kissler-Patig 2004) the contamination of
the globular cluster sample with unresolved background galaxies has
been discussed. Although we try to limit the contamination by setting
limits on $(V-H)$ and $(V-I)$ we can not entirely exclude it. These
color limits are derived from the SSP model isochrones, assuming a
maximum GC age of 15 Gyr and a maximum metallicity of
2.5$\times$Z$_{\odot}$ for the younger generation of GCs.
\end{itemize}

As already mentioned, both GCSs have been discussed
previously (e.g. \cite{woodworth00nic}, \cite{brodie00nic}) and to
further investigate our claims a direct comparison with previous
studies is useful. However, in the case of NGC~3311 this is only
possible in a more general way since the study by Brodie, Larsen \&
Kissler-Patig (2000), building on the same optical data, includes only
detections on the wide-field chips of HST/WFPC2, whereas combined
optical and near-infrared data are only available for the innermost
region, covered by the PC chip. 

\begin{figure*}[!ht]
\center
\includegraphics[width=6cm]{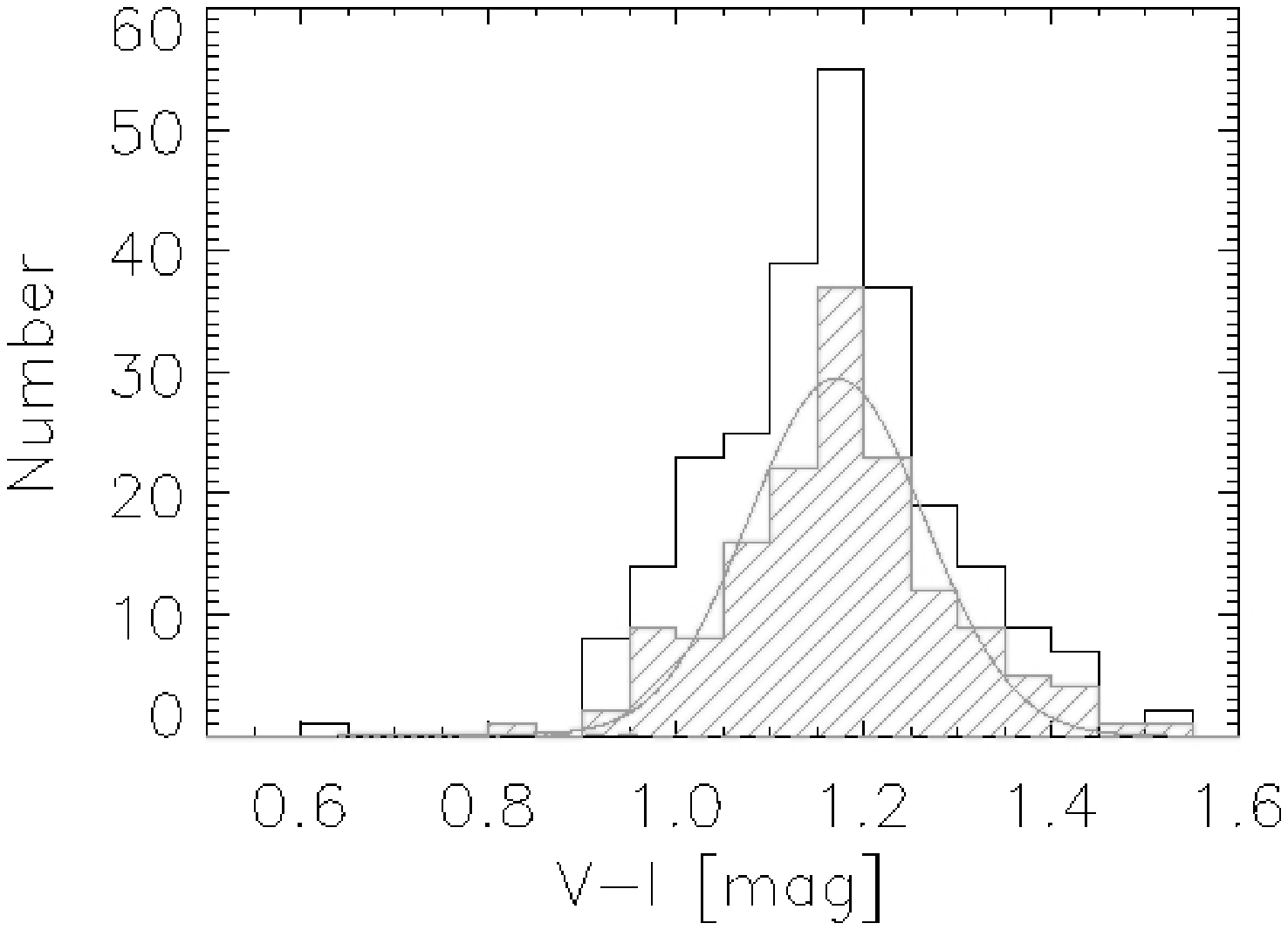}
\includegraphics[width=6cm]{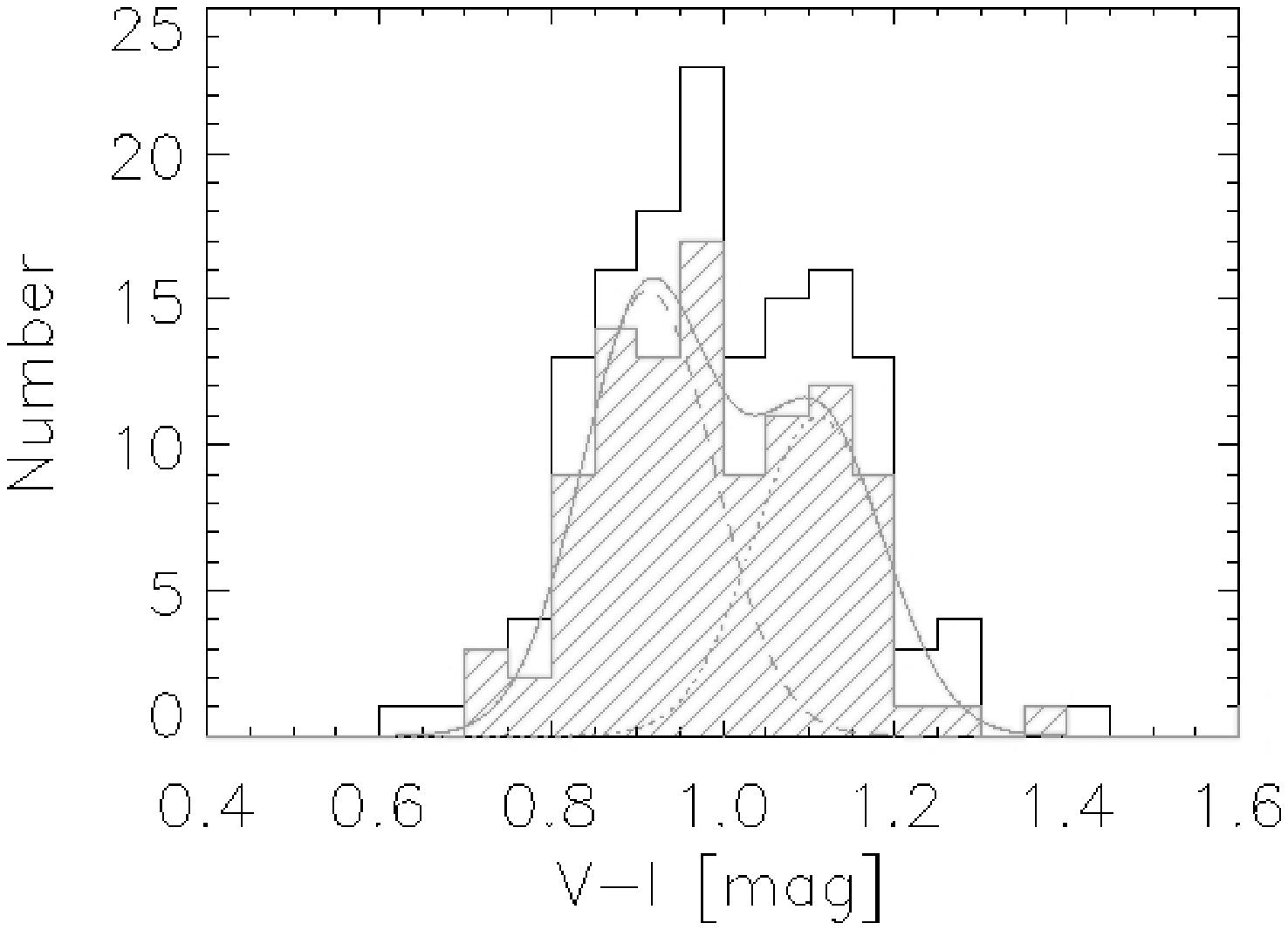}
\caption[VI distribution ]{$(V-I)$ color distribution for globular
clusters in IC~4051 (left) and NGC~3311 (right) detected on the PC
chip, for which as well H-band data are available. The open histogram
represents the complete VIH sample, whereas the shaded histogram
includes only objects for which the photometric errors,
$\delta$$(V-I)$ and $\delta$$(V-H)$ have been derived to be $<$0.15
mag. The solid curve shows either the single Gaussian (for IC~4051) or
the sum of the two Gaussian functions (dotted and dashed lines) in case of NGC~3311.}
\label{vidistribnic}
\end{figure*}

In IC~4051 the $(V-I)$ distribution for objects which are later also
detected in the infrared, is best fitted with a single Gaussian
centered on V-I=1.17$\pm$0.005 (see Figure \ref{vidistribnic}, left
panel), which is though somewhat redder , due to color selection, but still in agreement with
the results given by Woodworth \& Harris (2000) $(V-I)$=1.106. Using
the linear conversion by Woodworth \&
Harris (2000): $V-I$=0.17[Fe/H]+1.15 the mean $(V-I)$ colors translate
into mean metallicities of $\approx$+0.11 and $\approx$-0.3,
respectively. We note again that these values refer to different
samples, based on the observed field of view. 

In the discussion about the metallicity distribution for the NGC~3311
GCS we have to be especially cautious given the above caveats about
the dependence of the metallicity distribution on the derived ages and
is included rather for completeness than for solid
argumentation. Comparing the distribution of $(V-I)$ color, as shown
in Figure \ref{vidistribnic} (right panel), with the results obtained
for the wide-field chip detections, reveals as well a bimodal color
distribution with the center of the blue and red sub-population at
0.91$\pm$0.01~mag and 1.11$\pm$0.01~mag, respectively. In comparison,
in the work by Brodie, Larsen and Kissler-Patig (2000) the $(V-I)$
color distribution peaks at 0.91$\pm$0.03~mag and 1.09$\pm$0.03~mag,
resulting in a metallicity of [Fe/H]$\approx$-1.5 and $\approx$-0.75
dex. The color-metallicity conversion follows the relation given in
Kissler-Patig et al. (1998):
[Fe/H]=(-4.50$\pm$0.30)+(3.27$\pm$0.32)(V-I). In contrast, the
metallicities presented in this work, derived from the $(V-H)$ color
directly, but are based on the estimated age and a subsequent
application of the SSP model isochrones.

\vskip 0.5cm
\section{Summary}
\label{s:summarynic}
We have investigated the age structure of GCs in two luminous
early-type galaxies, IC~4051 and NGC~3311, each with large GC
populations, using the combination of optical and near IR colors to
help break the age-metallicity degeneracy. In both galaxies, we find
evidence for a significant fraction of young/intermediate age clusters
($\approx$2-~5 Gyr), in addition to their old counterparts, by
comparing the observed color distribution with models. This
result is confirmed by the comparison between the cumulative age
distribution in observed and simulated systems, which not only
indicates a second major cluster formation event but also allows to set
some constraints on the time frame. Further evidence for a less
passive evolution of the galaxies is given by independent features
(dust lanes, co-rotating core), but especially in NGC~3311, the
effects of small number statistics make our results tentative. Based
on the small number of objects in this sample, our detected fraction
of 50-- 60$\%$ of young globular clusters in the total GCS in NGC~3311
has a large uncertainty. The results for IC~4051 are more
statistically reliable, suggesting more than half of the globular
clusters are young/intermediate age objects. This fraction should be
regarded as an upper limit, considering that most of the GCs excluded
from the age analysis are probably mainly old, and that our
observations are biased to the central regions, where younger objects
should preferentially be found. Thus, at this stage the global
importance of the star forming event leading to a second generation of
globular clusters in NGC~3311 and IC~4051 cannot be firmly
established. More data, specially sprectroscopic, over a wider
galactocentric field, would be useful for clarifying the nature of
these populations. Nevertheless, the improved age resolution of combined
optical and near-infrared photometry reveals age sub-populations,
which could not be detected using optical data alone. 

Preceding the results presentet in Appendix \ref{appendixa} we want to
include the effects of SSP model dependency in this summary. Although
of little consequence for the age distribution the choice of the SSP
model is becomes important for the derived metallicity
distribution. The extremely high mean metallicity of the 2 Gyr old
population ( [Fe/H]$\approx$~0.8 dex, see Section \ref{s:metalnic}),
obtained with the Bruzual \& Charlot (2003) models, becomes much more
reasonable when the Maraston (2005) are applied,
i.e. [Fe/H]$\approx$~0.2. Whether this large discrepancy is based on
the models themselves, with only the Maraston models including the
TP-AGB phase, or is merely an artifact of the interpolation of the
metallicity, or both, still needs to be investigated. Nevertheless we
note that the difference in metallicity decreases with an inceasing
age of the cluster population. Due to the small number of metal-poor
objects we do not discuss model dependencies for this sub-sample
here. \\

\vskip 0.5cm
{\bf{Acknowledgments}} 

Support for this work was provided by NASA through grant number
GO-07280 from the Space Telescope Science Institute, which is operated
by the Association of Universities for Research in Astronomy, Inc.,
under NASA contract NAS5-26555. K.Mighell kindly provided advice
concerning the CCDCAP package. D.G. would like to thank Nick Suntzeff
and Malcolm Smith of the staff of Cerro Tololo Inter-American
Observatory for their kind support of this project. D.G. gratefully
acknowledges support from the Chilean {\sl Centro de Astrof\'\i sica}
FONDAP No. 15010003. The authors wish also to thank the anonymous referee for
her/his comments, which helped immensely to improve the manuscript. \\

\appendix
\section{SSP model dependency}
\label{appendixa}
\begin{figure*}[!ht]
\center
\includegraphics[width=8cm]{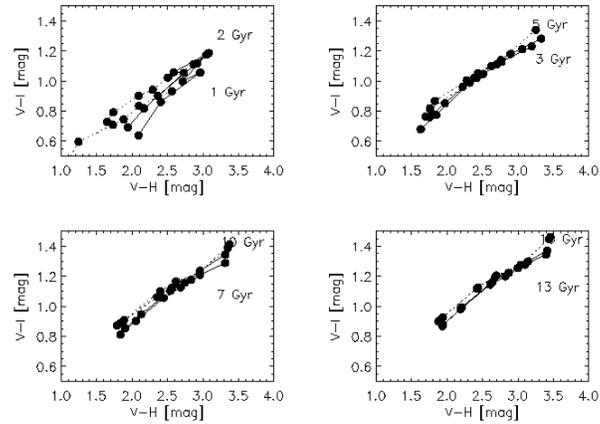}
\caption{Comparison between the color predictions of the Bruzual \&
Charlot (2003, dotted line) and the Maraston (2005,solid line) SSP models. Clearly the largest
discrepancy is found for objects younger than 3 Gyr, whereas for older
populations the difference in $(V-I)$ is in the range of the
photometric error or below. We note that
the models are given for different metallicities (Bruzual \& Charlot:
-2.25,-1.35,-0.64, -0.33,0.0,+0.56 dex; Maraston:-2.25,-1.35,-0.33,0.0,+0.35
dex).}
\label{compare_ma05_bc03}
\end{figure*}

In the recent past, tremendous progress has been made in the
understanding of stellar evolution and one of the major outcomes is more
and more accurate models of how colors and spectral features for
single stellar populations evolve in time, depending on metallicity and
other parameters, e.g. the structure of the horizontal branch
(e.g. \cite{maraston00nic}, \cite{lee02nic}). Although this paper does
not intend to argue in favor of a specific SSP model we would like to
show how the results depend on our choice of the SSP model. In the
main part of this paper we apply the new version of the Bruzual \&
Charlot model isochrones (2003). As shown in Hempel
\& Kissler-Patig (2004) and Hempel (2004) the choice between the Bruzual \& Charlot
models (2000, 2003) or, alternatively, the model by Vazdekis (1999) or
Maraston (2001) affect the results only marginally, i.e. the detection
of intermediate age clusters is essentially independent of the
selected model. However, in order to
investigate the subsequent advances made in SSP modeling we will as
well conduct our analysis applying the latest version of the SSP
models by Maraston (2005), assuming a red/intermediate horizontal
branch structure and a Salpeter IMF. The $(V-I)$ and $(V-H)$ colors
for both the Bruzual \& Charlot (2003) and Maraston (2005) model are
compared in Figure \ref{compare_ma05_bc03}. We note that unlike
earlier releases of both SSP models {\it{the color predictions for a given
age do not differ significantly, except for metal-poor objects with
ages $/leq$2 Gyr}}. Prior to our analysis we compared as well the model
predictions for the $(V-I)$ and $(V-H)$ color assuming a
{\bf{red/intermediate}} and a {\bf{blue}} horizontal branch and
various ages ($>$ 1Gyr) as offered by the SSP models. With the
exception of a 1 Gyr old stellar population, the change in color is
smaller than the photometric error in the observational data and in
fact smaller than the color difference between models using different
IMF's, especially within the red $(V-H)$ color range where our
technique is sensitive to the detection of young /intermediate GCs.\\

\begin{figure*}[!ht]
\center
\includegraphics[width=8cm]{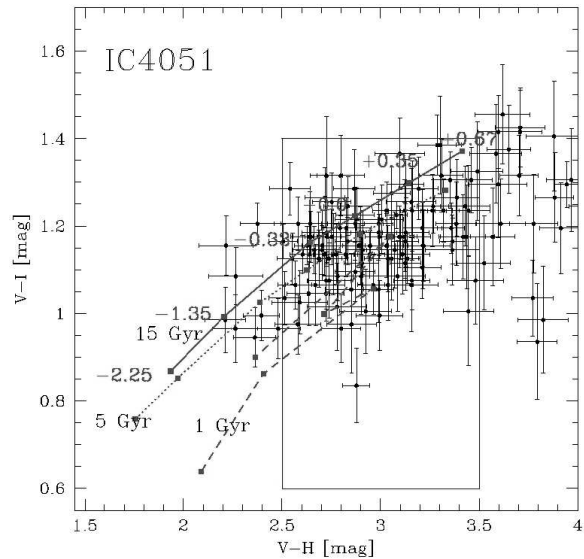}
\caption{$(V-I)$~$vs.$~$(V-H)$ color-color diagram for IC~4051. As in
Figure \ref{ic4051f6nic} only GCs with photometric errors $<$0.15 mag
have been selected. The 1,5, and 15 Gyr isochrones are marked as solid,
dotted and dashed line, respectively.Here the SSP model by Maraston
has been applied.}
\label{ic4051_ma05}
\end{figure*}

\begin{figure*}[!ht]
\center
\includegraphics[width=8cm]{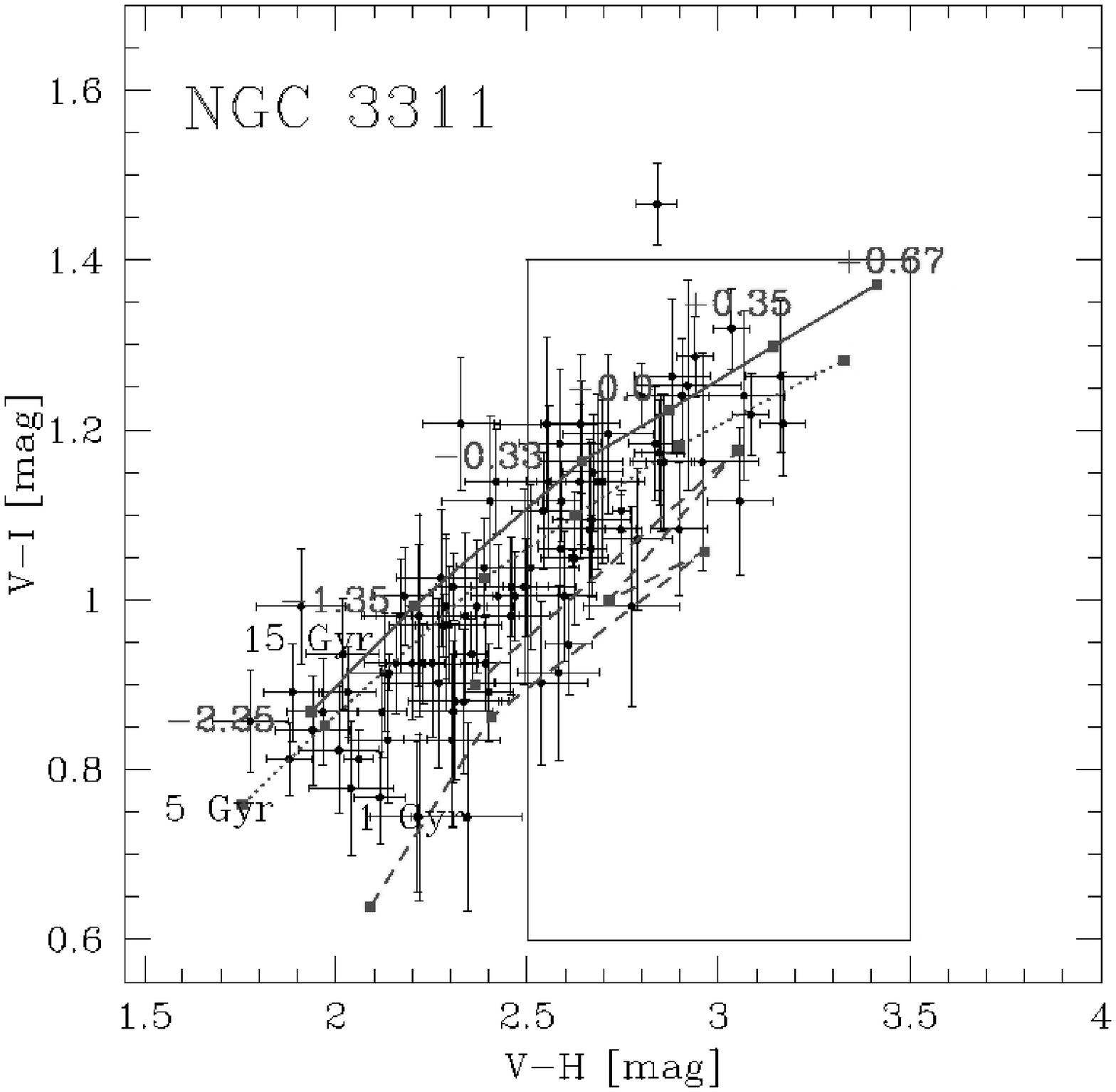}
\caption{$(V-I)$~$vs.$~$(V-H)$ color-color diagram for NGC~3311. As in 
Figure \ref{ic4051f6nic} only GCs with photometric errors $<$0.15 mag
have been selected. The 1,5, and 15 Gyr isochrones are marked as solid,
dotted and dashed line, respectively. Here the SSP model by Maraston
has been applied.}
\label{n3311_ma05}
\end{figure*}

The color-color diagrams, for both globular cluster systems are again
shown in Figures \ref{ic4051_ma05} and \ref{n3311_ma05}. In contrast
to Figure \ref{ic4051f6nic} and \ref{n3311colcolnic} the colors are
now compared to the SSP models by Maraston (2005). Still the color
distributions show a large fraction of globular clusters with a
$(V-I)$ color which is too blue to be consistent with a population
older than 5 Gyr.

In order to detect GC sub-populations within a galaxy we derive the
cumulative age distribution (see \cite{hempel03nic},
\cite{hempel04anic}) based on optical and near-infrared colors in the same way as in Section 3.2. Hereby
the GC density with respect to the model isochrones is of more
importance than the color of each GC separately. As described before,
the GC density is determined as the number of objects with a $(V-I)$
color redder and therefore older than a given isochrone. Previously
those number counts were estimated for the 1,2,3,5,7,10,13 and 15 Gyr
isochrone. Using the new Maraston models requires a small alteration
in this procedure. Despite the non-monotonic pattern of the 1 Gyr
isochrone we follow the same counting procedure, taking advantage of
the fact that, independent from metallicity, the model predictions for
$(V-I)$ give a redder color for a 2 Gyr old population compared to 1
Gyr population. Nevertheless, in the simulations we assume the second
generation of GC to be at least 2 Gyr old. \\

\begin{figure*}[!ht]
\center
\includegraphics[width=8cm]{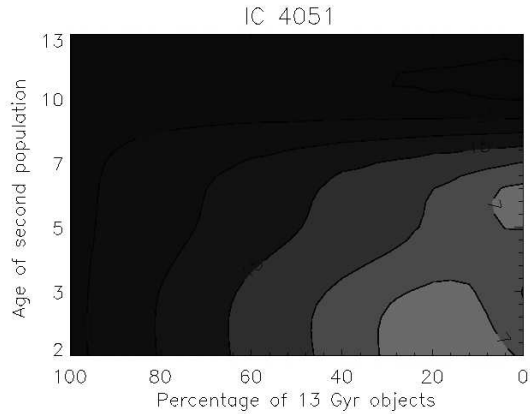}
\caption{$\chi$$^2$-test result for IC~4051. Different levels
represent the $\chi$$^2$-values of the comparison between the cumulative age
distribution in the selected GC sample and a set of 66 simulated
systems. For both, the modelling and the determination of the age
distribution the SSP models by Maraston (2005) have been
applied. Similar to the application of the Bruzual \& Charlot models
(2000) the best fitting models contain a large fraction of
young/intermediate age globular clusters.}
\label{contouric4051ma05}
\end{figure*}

\begin{figure*}[!ht]
\center
\includegraphics[width=8cm]{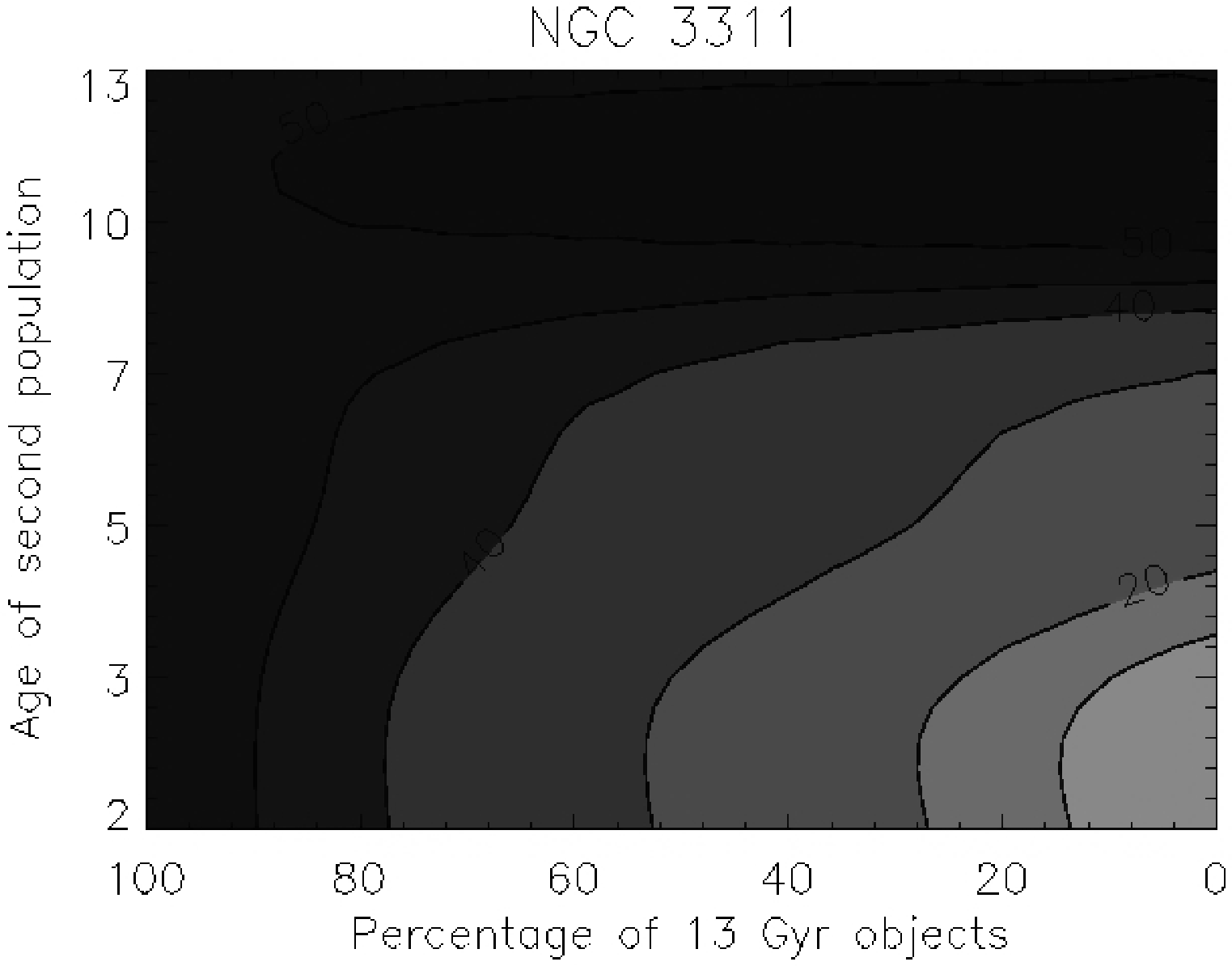}
\caption{Contour plot representing the results of the $\chi$$^2$-test
comparing the cumulative age distribution in the NGC~3311 GCS
(selected sample) and the Monte-Carlo simulations. As in Figure
\ref{contouric4051ma05} the SSP models by Maraston (2005) have been used.}
\label{contourn3311ma05}
\end{figure*}

Comparing the cumulative age distributions in both observed systems
with our set of simulations, shows at first significant
differences, if the Maraston SSP models are applied (Figures
\ref{contouric4051ma05} and \ref{contourn3311ma05}). Hence we need to
emphasize again that this approach does not aim at high precision age
estimates but is rather a tool to detect significant sub-populations
of GCs. Ideally, a successful detection should be complemented by
spectroscopic investigations to determine the age more accurately. Due
to the large distance of most early-type galaxies (especially in high
density environment) spectroscopic surveys are hardly feasible.
Keeping this in mind we find that the conclusions about a second
population of intermediate age globular clusters with the selected
sample is confirmed, although the relative size of the second GC
generation is shifted to larger values, i.e. using the Maraston 2005
models leads to an even larger fraction of intermediate age
clusters. Since the latter is affected by various biases (color cuts,
spatial bias, etc.) we consider large numbers of intermediate/young
GCs to be an upper limit at best. Considering the importance of
horizontal branch structure for the modeling of SSP isochrones this
result seems to be surprising. On the other side we have to be aware
that a major impact of the HB structure is expected for the colors of
metal-poor objects (see Figure\ref{compare_ma05_bc03}), which we
exclude from our analysis by setting color cuts.

To round things up we will finish this Section by deriving the
metallicity distribution in IC~4051, following the procedure described
in Section
\ref{s:metalnic}, but now based on the Maraston (2005) models. 
As before we calculate the metallicity under the assumption of the
globular clusters being 2,~5, or 13 Gyr old, depending on their
$(V-I)$ color with respect to the corresponding isochrones. The metallicity [Fe/H]
for each objects is then interpolated for a given
$(V-H)$. \\

\begin{figure*}[!ht]
\center
\includegraphics[width=12cm]{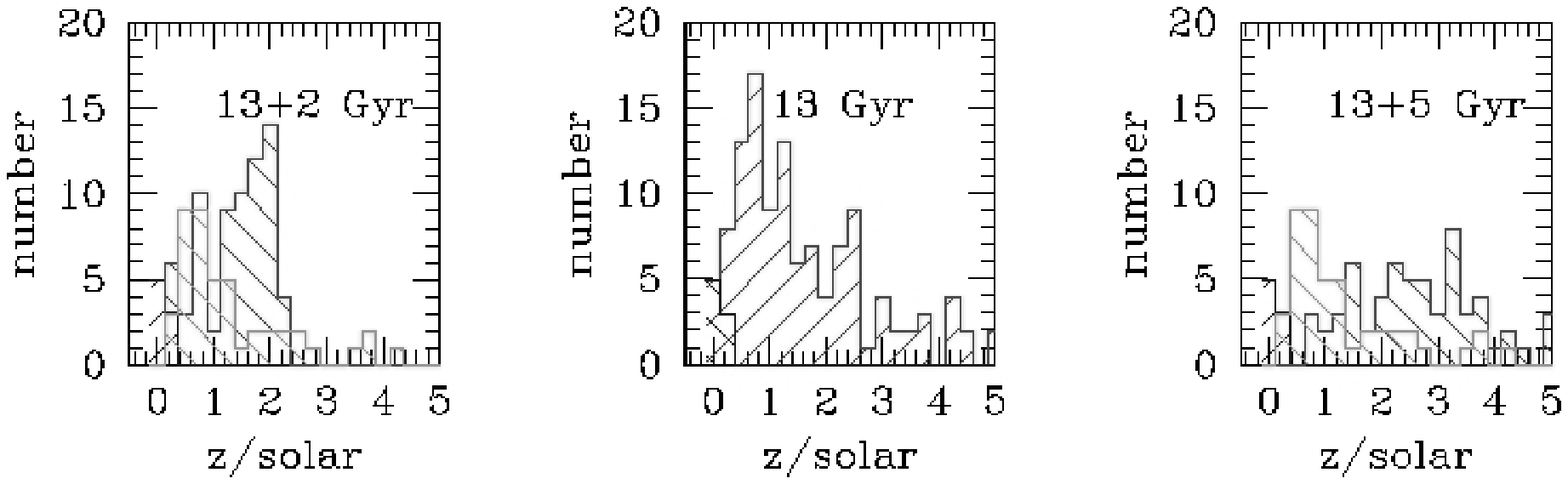}
\includegraphics[width=12cm]{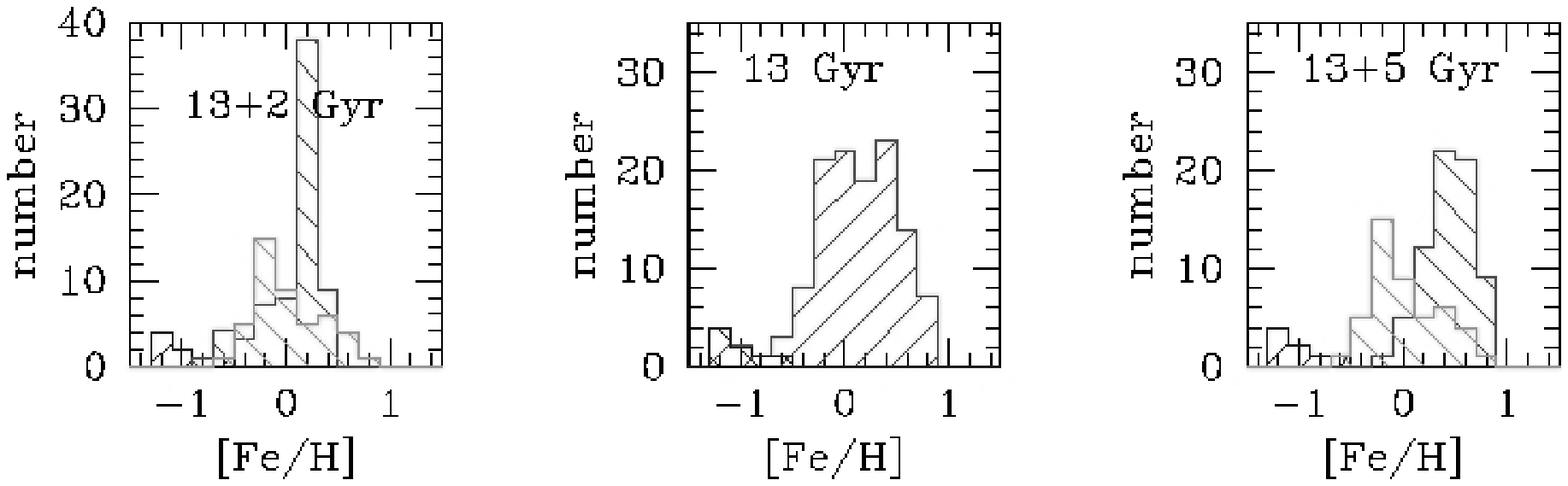}
\caption[Metallicity distribution in IC~4051]{Metallicity distribution
in IC~4051 assuming different GC age distributions. We show the
metallicity distribution both in terms of linear metal abundance
(upper panel) as well as in logarithmic metallicity [Fe/H] (lower
panels). The different panels and the color coding of the histograms correspond to Figure \ref{metal_ic4051nic}. }
\label{metal_ic4051nic_ma05}
\end{figure*}

Comparing the color predictions of both models we note that stellar
populations older than 3 Gyr are almost unaffected, the Bruzual \&
Charlot $(V-I)$ colors being somewhat bluer than their Maraston
counterparts. Translated into ages the globular clusters are slightly
younger when the Maraston models are applied. There is, however, a
significant difference in $(V-H)$ with respect to the corresponding
metallicity values. As shown in Figure \ref{compare_ma05_bc03} the
$(V-H)$ color of a 1 or 2 Gyr old population of low metallicity
(i.e. [Fe/H]$\approx$-2.3) is much bluer in the Bruzual \& Charlot
models. Consequently, the metallicity distribution for the IC~4051
GCs will be shifted toward lower metallicities when the Maraston
models are applied. This can be seen in the left panels of Figure
\ref{metal_ic4051nic_ma05}. The metallicity for 2 Gyr old objects (red
histogram) peaks at much lower values compared to Figure
\ref{metal_ic4051nic}, left panels. For the first generation of
GCs (13 Gyr) the metallicity estimates change only marginally and are
also similar for a 5 Gyr population within the anticipated accuracy
of 0.2 dex.

Based on the above results we therefore conclude that our primary
goal- the detection of age sub-populations in GCSs- is not
significantly influenced by the choice of the SSP model. Nevertheless,
we note that our method is somewhat handicapped when the SSP model
isochrones are not monotonic, e.g. for a given age an unique correlation
between the two colors can not be established. There is, however, no doubt
that state of the art SSP models should be applied, especially when
the information we want to extract from the globular cluster colors
become more and more detailed (including metallicity distributions)
and an increasing accuracy of the data allows more precise
calculations.


\begin{thebibliography}{} 
\bibitem[Ashman \& Zepf~1998]{ashman98nic}Ashman, D.M. \& Zepf,
 S.E.~1998,''Globular Cluster Systems'', Cambridge University Press
\bibitem[Barmby et al.~2000]{barmby00nic}Barmby, P., Huchra,
 J.P., Brodie, J.P., et al.~2000,~\aj,~119,~727
\bibitem[Baum et al.~1997]{baum97nic}Baum, W.A., Hammergren, M. Thomsen,
B. et al. 1997, \aj, 113, 1483
\bibitem[Bertin \& Arnout~1996]{bertin96nic}Bertin, E., Arnout,
S.~1996,~A\&AS,~117,~393B
\bibitem[Brodie, Larsen, \& Kissler-Patig~2000]{brodie00nic}Brodie, J.P.,
Larsen, S., Kissler-Patig, M.~2000,~\apj,~543,~L19
\bibitem[Brodie et al. 2005]{brodie05nic}Brodie, J.P., Strader, J.,
Denicol\'o, G. et al.~2005, \aj, in press (astro-ph/0502467)
\bibitem[Bruzual \& Charlot~1993]{bruz93nic}Bruzual, G.A.\& Charlot,
S.~1993, \aj,~405,538
\bibitem[Bruzual~2000]{bruz00nic}Bruzual, G.A.~2000, private communication
\bibitem[Bruzual \& Charlot~2003]{bruz03nic}Bruzual, G.A.\& Charlot,
S.~2003, \mnras, 344,~1000
\bibitem[Cohen \& Matthews~1994]{cohen94nic}Cohen, J.G. \& Matthew,
K.~1994, \aj, 108, 128
\bibitem[Davies et al.~2001]{davies01nic}Davies, R.L.,
    Kuntschner, H., Emsellem, E. et al. ~2001, \aj,~548,~33
\bibitem[deVaucouleur et al.~1991]{rc3nic}deVaucoleurs, G., deVaucouleurs,
A., Corwin, H.G.et al.~1991, ``Third Ref.Catalogue of Bright
Galaxies'', Springer New York
\bibitem[Dickinson et al.~2002]{dickinson02nic}Dickinson, M., Sosey,
M., Rieke, M. et al.~2002, "NICMOS Photometric Calibration", in the
proceedings of the ``2002 HST Calibration Workshop'', Space Telescope
Science Institute, Baltimore, Maryland, 17./18.10. 2002, eds. Santiago
Arribas, Anton Koekemoer, and Brad Whitmore,~233
\bibitem[Ferrari et al.~1999]{ferrari99nic}Ferrari, F., Pastoriza,
 M.G., Macchetto, F. et al.  1999, A\&ASS, 136, 269
\bibitem[Gebhardt \& Kissler-Patig~1999]{gebhardt99nic}Gebhardt, K., \&
Kissler-Patig, M.~1999,~\aj,~118,~1526
\bibitem[Geisler, Lee \& Kim~1996]{geisler96nic}Geisler, D., Lee, M.G.,
\& Kim, E.~1996, \aj,~111, 1529
\bibitem[Goudfrooij et al.~2001a]{goudfrooij01anic}Goudfrooij, P.,
Mack, J., Kissler-Patig, M. et al.~2001, \mnras,~322,~643
\bibitem[Goudfrooij et al.~2001b]{goudfrooij01bnic}Goudfrooij, P.,
Alonso, M.V., Maraston, C. et al.~2001, \mnras,~328,~237
\bibitem[Grillmair et al.~1994]{grillmair94nic}
Grillmair, C.J., Faber, S.M., Lauer, T.R. et al.~(1994),~\aj,~108, 102
\bibitem[Harris \& van den Bergh~1981]{harris81nic}Harris, W.E.,\& van
den Bergh, S.~1981, \aj,~85, 1627
\bibitem[Harris, Smith, \& Myra~1983]{harris83nic}Harris, W.E., Smith,
M.G., Myra, E.S.~1983, \apj,~272,~456
\bibitem[Harris~1991]{harris91nic}Harris, W.E.~1991, \araa,~29,~543
\bibitem[Harris~2001]{harris01nic}Harris, W.E.~2001, in Star Clusters, Saas-Fee
Advanced Course 28, Swiss Society for Astrophysics and Astronomy, ed. L.Labhardt
\& B.Binggeli (Berlin: Springer-Verlag), 223
\bibitem[Hempel et al.~2003]{hempel03nic}Hempel, M., Hilker, M.,
Kissler-Patig, M. et al.~2003, \aap,~405,~487, Paper III
\bibitem[Hempel \& Kissler-Patig 2004]{hempel04anic}Hempel, M. \&
Kissler-Patig, M.~2004, \aap,~419,~863, Paper IV
\bibitem[Hempel~2004]{hempel04bnic}Hempel, M. ~2004, PhD thesis,
``Early-Type Galaxies and Their Sometimes Not So Old Globular
Clusters'', Ludwig-Maximilian's-University Munich
\bibitem[Hibbard \& Mihos~1995]{hibbard95nic}Hibbard, J.E. \& Mihos,
J.C.~1995,~\aj,~110,~140
\bibitem[Hilker~2002]{hilker02nic}Hilker, M.~2002,
In: Proceedings of the IAU Symp. 207 ``Extragalactic Star Clusters'',
eds. D.~Geisler, E.K.~Grebel, \& D.~Minniti; San Francisco: ASP,~281
\bibitem[Hilker~2003a]{hilker03anic}Hilker, M.~2003, in ``New Horizons
in Globular Cluster Astronomy'', eds.  G.~Piotto, G.~Meylan,
G.~Djorgovski, \& M.~Riello, ASP Conf. Ser., 296,~583
\bibitem[Hilker~2003b]{hilker03bnic}Hilker, M.~2003,
in ESO Astrophysics Symposia, ``Extragalactic Globular Cluster
Systems'', ed. M.~Kissler-Patig, Springer,~173
\bibitem[Holtzman et al.~1995]{holtzman95nic}Holtzman, J.A., Burrows,
C.J., Casertano, S. et al.~1995, \aj,~107,~1065
\bibitem[Jensen, Tonry \& Thomson~(2001)]{jensen01nic}Jensen, J.B., Tonry,
J.L., Thomson, R.I. et al.~2001,~\apj,~550,~503
\bibitem[Jorgensen \& Franx~1994]{jorgensen94nic}Jorgensen, I. \&
Franx, M.~1994, \apj, 433,~553
\bibitem[Jura~1986]{jura86nic}Jura, M. ~1986, \apj,~306, 483
\bibitem[Kent \& Gunn~1982]{kent82nic}Kent, S.M. \& Gunn,
J.E..~1982, \aj, 87, 945
\bibitem[Kissler-Patig et al.~1998]{kissler98nin}Kissler-Patig,
M., Brodie, J.P., Schroder, L.L. et al.~1998, \aj, 115, 105
\bibitem[Kissler-Patig~2002]{kisslerpatig02nic}Kissler-Patig, M. 2002, IAU Symp.207, 207
\bibitem[Kissler-Patig et al.~2002]{kissler02nic}Kissler-Patig, M.,
Brodie, J.B., \& Minniti, D.~2002, \aap, 391, 441, Paper I
\bibitem[Kundu \& Whitmore~2001a]{kundu01anic}Kundu, A.\& Whitmore,
B.C.~2001, \aj,~121,2950
\bibitem[Kundu \& Whitmore~2001b]{kundu01bnic}Kundu, A.\& Whitmore,
B.C.~2001, \aj,~122, 1251
\bibitem[Kuntschner, Smith \& Colless~2002]{kuntschner02nic}Kuntschner,
H., Smith, R.J., Colless, M. et al.~2002,~\mnras,~337,~172
\bibitem[Larsen et al.~2003]{larsen03nic}Larsen, S., Brodie, J.P.,
Beasley, M.A. et al.~2003, \apj, 585, 767
\bibitem[Lee, Lee \& Gibson~2002]{lee02nic}Lee, H.-C., Lee,
Y.-W., Gibson, B.K.~2002, \aap, 124, 2664
\bibitem[Li, Mac Low \& Klessen~2004]{li04nic}Li, Y., Mac Low,
M.-M., Klessen, R.S.~2004, \apj, submitted, (astro-ph/0407248)
\bibitem[Malhotra~2002]{malhotra02nic}Malhotra, S. ~2002, ``NICMOS
Instrument Handbook'', Version 5.0 (Baltimore: STScI)
\bibitem[Maraston \& Thomas~2000]{maraston00nic}Maraston, C. \&
Thomas, D.~2000, \apj, 541, 126
\bibitem[Maraston, Greggio \& Thomas~2001a]{maraston01nic} Maratson,
C., Greggio, L., Thomas, D.~2001, Ap\&SS, 276, 893
\bibitem[Maraston et al. 2001b]{maraston01bnic}Maraston, C.,
Kissler-Patig, M., Brodie, J.P. et al. ~2001, \aap, 370,176
\bibitem[Maraston~2005]{maraston05nic}Maraston, C.~2005,\mnras, submitted
\bibitem[McLaughlin et al.~1995]{mclaughlin95nic}McLaughlin, D.E., Secker, J.,
Harris, W.E et al.~1995, \aj,~118,~1033
\bibitem[McNamara, Bregman \& O'Connell~1990]{mcnamara90nic}McNamara,
B.R., Bregman, J.N.,\& O'Connell, R.W.~1990, \aj,~360,~20
\bibitem[Mehlert et al.~1998]{mehlert98nic}Mehlert, D., Saglia,
R. P., Bender, R. et al.~1998,~\aap,~332,~33
\bibitem[Mehlert et al. 2003]{mehlert03nic}Mehlert, D., Thomas,
D., Saglia, R.P. et al. 2003,~\aap, 407, 423
\bibitem[Mighell \& Rich~1995]{mighell95nic}Mighell, K.J., Rich, R.M.~1995,~\aj,~110,~1649
\bibitem[Minniti et al.~1996]{minniti96nic}Minniti, D., Alonso, M.V.,
Goudfrooij, P. et al.~1996,~\apj,~467, ~221
\bibitem[Puzia et al.~2002]{puzia02nic}Puzia,T.H., Zepf, S.E.,
Kissler-Patig, M.et al.~2002, \aap, 391, 453, Paper II
\bibitem[Puzia et al.~2004a]{puzia04anic}Puzia, T.H., Kissler-Patig,
M., Thomas, D. et al.~2004, \aap, 415, 123
\bibitem[Puzia et al.~2004b]{puzia04bnic}Puzia, T.H., Kissler-Patig, M.,
Thomas, D. et al.~2004, \aap, submitted
\bibitem[Renzini~1981]{renzini81nic}Renzini, A.~1981, \aj, 94, 175
\bibitem[Richter~1989]{richter89nic}Richter, O.-G.~1989,~\aaps, 77, 237
\bibitem[Searle \& Zinn~1978]{searle78nic}Searle, L. \& Zinn, R. \apj,
225, 357
\bibitem[Secker et al.~1995]{secker95nic}Secker, J., Geisler, D.,
McLaughlin, D.E. et al. 1995, \aj, 109, 1019
\bibitem[Schlegel, Finkbeiner \& Davis~1998]{schlegel98nic}Schlegel,
D.A., Finkbeiner, D.P., Davis, M.~1998, \apj,~500,525
\bibitem[Schweizer \& Seitzer~1998]{schweizer98nic}Schweizer, F. \&
Seitzer P.~1998,~\aj,~116,~2206
\bibitem[Stephens et al.~2000]{stephens00nic}Stephens, A.W., Frogel,
J.A., Ortolani, S, et al.~2000,\aj,~119,419
\bibitem[Thomas, Maraston, \& Bender~2003]{thomas03anic}Thomas, D.,
Maraston, C.,\& Bender, R.~2003, \mnras, 339, 897
\bibitem[Thomas \& Maraston~2003]{thomas03bnic}Thomas, D., \&
Maratson, C.~2003,~\aap,~401,~429
\bibitem[Tully~1988]{tully88nic}Tully, B.~1988, ``Nearby galaxies
catalog'', Cambridge University Press
\bibitem[V\"asterberg, J\"ors\"ater \&
Lindblad~1991]{vasterberg91nic}V\"asterberg, A.R., J\"ors\"ater, S.,
\& Lindblad, P.O.~1991, \aap, 247, 335
\bibitem[Vazdekis~1999]{vaz99nic}Vazdekis, A.~1999, \apj, 513, 224
\bibitem[Vazdekis et al.~2003]{vaz03nic}Vazdekis, A., Cenarro, A.J.,
Gorgas, J. et al.~2003,~\mnras,~340, 1317
\bibitem[Whitmore \& Schweizer~1995]{whitmore95nic}Whitmore, B.C. \&
Schweizer, F.~1995, \aj,
\bibitem[Wielen~1990]{wielen90nic}Wielen, R. (ed.)~1990, Dynamics and
Interactions of Galaxies, Springer Verlag
\bibitem[Woodworth \& Harris~2000]{woodworth00nic}Woodworth, S.C.,
Harris, W.E.~2000,~\aj,~119,~2699
\bibitem[Worthey~1994]{worthey94nic}Worthey, G.~1994, \apjs,~95,~107
\bibitem[de Zeeuw et al. 2002]{zeeuw02nic}de Zeeuw, P.T., Bureau, M.,
    Emsellem, E. et al.~2002, \mnras, 329, 513
\bibitem[Zepf \& Ashman~1993]{zepf93nic}Zepf, S.~\& Ashman, D.~1993,
\mnras, 264, 611

\end{thebibliography}
\end{document}